\pdfoutput=1

\documentclass[a4paper,11pt]{article}

\usepackage{jcappub} 

\usepackage[T1]{fontenc} 
\usepackage{caption}
\usepackage{subcaption}
\usepackage{graphicx}
\usepackage{mathrsfs}
\usepackage{amsmath}
\usepackage{amssymb}
\usepackage{mathtools} 
\usepackage{soul}

\providecommand{\abs}[1]{\lvert#1\rvert}
\providecommand{\bd}[1]{\boldsymbol{#1}}
\providecommand{\ro}[1]{\mathrm{#1}}

\providecommand{\Mp}{M_{\mathrm{Pl}}}

\setstcolor{blue}
\sethlcolor{cyan}

\title{\boldmath Parker Bounds on Monopoles with Arbitrary Charge from Galactic and Primordial Magnetic Fields}

\author{Takeshi Kobayashi}
\author{and Daniele Perri}

\affiliation{SISSA, International School for Advanced Studies, \\via Bonomea 265, 34136 Trieste, Italy\\}

\affiliation{INFN, Sezione di Trieste, \\via Valerio 2, 34127 Trieste, Italy\\}

\affiliation{IFPU, Institute for Fundamental Physics of the Universe, \\via Beirut 2, 34014 Trieste, Italy\\}

\emailAdd{takeshi.kobayashi@sissa.it}
\emailAdd{dperri@sissa.it}

\abstract{
We present a comprehensive study of Parker-type bounds on magnetic monopoles with arbitrary magnetic charge, including minicharged monopoles and magnetic black holes. 
We derive the bounds based on the survival of galactic magnetic fields, seed magnetic fields, as well as primordial magnetic fields.
We find that monopoles with different magnetic charges are best constrained by different astrophysical systems:
while monopoles with a Dirac charge are tightly constrained by seed galactic magnetic fields, minicharged monopoles are strongly constrained by primordial magnetic fields, and magnetic black holes by the density of dark matter.
We also assess the viability of the various types of monopoles as dark matter, by studying whether they can cluster with galaxies hosting magnetic fields.
}

\begin{document}
\maketitle
\flushbottom

\section{Introduction}
\label{sec:intro}

Magnetic monopoles have long been a topic of intense study since Dirac showed that their existence is consistent with quantum electrodynamics~\cite{Dirac:1931kp}.
This discovery was followed by 't~Hooft~\cite{tHooft:1974kcl} and Polyakov~\cite{Polyakov:1974ek} who found classical soliton solutions that correspond to monopoles.
Such solitonic monopoles can be produced during phase transitions in the early universe~\cite{Preskill:1979zi,Zeldovich:1978wj,Vilenkin:2000jqa} and are an inevitable prediction of theories of grand unification. 
The magnetic charge of monopoles is constrained by the Dirac quantization condition as $e g= 2 \pi n $, $n \in \mathbb{Z}$.
For this reason, experimental searches over the years have mostly focused on monopoles with a charge $g \sim 2 \pi / e$.
However recently a number of theoretical works have considered monopoles possessing a wide range of charges.

Minicharged monopoles with $ g \ll 2 \pi / e$ can be realized by having a physical Dirac string.
Such configurations can arise, for instance, from a kinetic mixing between the Standard Model photon and a dark massive photon, in which case the monopole's
charge under the visible magnetic fields is proportional to the mixing parameter~\cite{Brummer:2009cs,DelZotto:2016fju,GomezSanchez:2011orv,Hook:2017vyc,Hiramatsu:2021kvu,Graesser:2021vkr}. 
Going to very large masses, magnetically charged black holes can be seen as giant monopoles with small charge-to-mass ratio.
The phenomenology of black holes with magnetic charge has recently been discussed in \cite{Araya:2022few,Maldacena:2020skw, Bai:2020spd,Zhang:2023tfv,Ghosh:2020tdu,Wadekar:2022ymq,Diamond:2021scl,Araya:2020tds}.
Such black holes are interesting as they cannot Hawking evaporate beyond extremality, leading to the possibility for primordial black holes with very small masses to survive until today.
Both minicharged monopoles and magnetic black holes have also been considered as interesting candidates of dark matter.

The relic abundance of magnetic monopoles is constrained by the requirement that they do not exceed the critical density of the universe \cite{ParticleDataGroup:2020ssz, Preskill:1979zi, Kolb:1990vq}.
However, even stronger constraints can be obtained from the magnetic fields present in the universe.
The idea behind this is that magnetic fields lose energy by accelerating monopoles, hence requiring their survival imposes an upper bound on the monopole abundance.
This was first proposed by Parker, who derived an upper bound on the monopole flux inside our Galaxy from the survival of the Galactic magnetic fields~\cite{Turner:1982ag, Parker:1970xv, Parker:1987}.
This so-called Parker bound was subsequently extended by considering a seed magnetic field of our Galaxy \cite{Adams:1993fj}. 
Intergalactic magnetic fields~\cite{Tavecchio:2010mk,Neronov:2010gir,Dermer:2010mm,MAGIC:2022piy}, on the other hand, may not directly yield Parker-type bounds. This is because the accelerated monopoles do not effectively dissipate their kinetic energy in the intergalactic voids, and thus can end up returning the energy to the magnetic field. However, if the intergalactic fields have a primordial origin, as suggested by various studies (see e.g. \cite{Subramanian:2015lua} for a review), the monopoles could have shorted out the magnetic fields in the early universe by transferring the magnetic energy into the cosmic plasma.
Parker-type bounds from primordial magnetic fields have thus been derived based on the fields' survival during the radiation-dominated epoch~\cite{Long:2015cza} and the reheating epoch~\cite{Kobayashi:2022qpl}.
We also note that strong magnetic fields can give rise to monopole pair production through the magnetic dual of the Schwinger effect~\cite{Schwinger:1951nm,Affleck:1981ag,Affleck:1981bma}.
Lower bounds on the monopole mass have been obtained by analyzing this effect on the surface of magnetars \cite{Gould:2017zwi,Hook:2017vyc}, in heavy-ion collisions at the LHC \cite{Gould:2017zwi,MoEDAL:2021vix}, and in primordial magnetic fields \cite{Kobayashi:2021des,Kobayashi:2022qpl}.

Direct searches for monopoles mainly rely on the detection of an induced electric current in superconducting rings \cite{Cabrera:1982gz}, or of the energy released into calorimeters from the interactions of a crossing monopole with the charged particles of the material \cite{MACRO:2002jdv,IceCube:2021eye}. 
However, it is extremely difficult to apply these methods to minicharged monopoles due to the sensitivity of the detectors and the selecting algorithms used in the experiments. 
For magnetic black holes, their very large masses combined with the constraint from the critical density of the universe restrict their flux on Earth to be extremely tiny; hence they are also minimally constrained by direct searches.
We should note that a subclass of GUT monopoles can catalyze nucleon decay, and searches based on this process have been performed; however whether monopole catalysis happens depends on the details of the model.
Thus, the possibility of deriving indirect bounds from astrophysical observations is even more compelling for monopoles possessing charges that are very different from the Dirac charge.

In this work we present a comprehensive study of Parker-type bounds on the flux of magnetic monopoles with arbitrary charge, including minicharged monopoles and magnetically charged black holes. 
We derive the flux bounds based on the survival of galactic magnetic fields, seed magnetic fields, and primordial magnetic fields, by clarifying the range of applicability of each bound along the way.
We find that, depending on the type of monopoles, the strongest bound arise from different astrophysical systems.
In particular, we show that while seed galactic magnetic fields impose tight bounds on monopoles with a Dirac charge, minicharged monopoles are strongly constrained by primordial magnetic fields, and magnetic black holes by comparison with the dark matter density.
We also derive conditions for monopoles to be able to cluster with galaxies hosting magnetic fields, based on which we examine whether the various types of monopoles can provide viable dark matter candidates.

This paper is organized as follows. In Section~\ref{sec:Parker} we revisit bounds from galactic fields and extend them to monopoles with arbitrary charge.
In Section~\ref{sec:PrimParker} we review the evolution of primordial magnetic fields in the presence of monopoles and derive bounds based on the survival of primordial fields. 
In Section~\ref{sec:comparison} we make a comparison of the different Parker bounds.
In Section~\ref{sec:BH} we investigate how the bounds apply to extremal magnetic black holes. We then conclude in Section~\ref{sec:concl}.
Appendix~\ref{app:dynamics} is dedicated to a study of monopole dynamics in galactic magnetic fields.

Throughout this work we use Heaviside-Lorentz units, with $c = \hbar = k_B = 1$, and use $\Mp$ to denote the reduced Planck mass~$ (8 \pi G)^{-1/2}$.
We denote the monopole's mass by~$m$, and the amplitude of the magnetic charge by~$g$.
The charge of a Dirac monopole is written as $g_{\ro{D}} = 2 \pi / e \approx 21$.

\section{Bounds from galactic magnetic fields}
\label{sec:Parker}

In this section we revisit the Parker bounds on the monopole flux from galactic magnetic fields~\cite{Parker:1970xv,Turner:1982ag} and seed fields~\cite{Adams:1993fj}. We extend the previous computations to allow for the monopoles to carry arbitrary magnetic charge,
and we also clarify the range of applicability of the bounds.

Let us consider a generic galaxy hosting magnetic fields, that are amplified by dynamo action with a time scale~$\tau_{\ro{gen}}$.
After the dynamo saturates, the magnetic field is assumed to stay nearly constant, and we represent the time period between saturation and today by~$\tau_{\ro{sat}}$. 
All cases with $\tau_{\ro{sat}}$ being comparable to or smaller than~$ \tau_{\ro{gen}}$ describe a similar situation where the fields have been growing until very recent times.
Hence, without loss of generality we impose $\tau_{\ro{sat}} \geq \tau_{\ro{gen}}$.

Monopoles within a galaxy are accelerated by the magnetic fields.
We model the fields such that they exist in a region of size~$R$, which is further divided into cells of uniform field. 
The size of each cell, i.e. the magnetic field's coherence length, is denoted by $l_{\ro{c}}$ ($ < R$). We further assume that the field strength~$B$ is the same in all cells, but the direction of the field is uncorrelated from one cell to the next. The average energy gain per monopole after it has passed through $N$ uncorrelated cells is derived in Appendix~\ref{app:dynamics} as
\begin{equation}
\Delta E_N \sim
  \begin{dcases}
\frac{N}{4} 
\frac{(g B l_{\ro{c}})^2}{m (\gamma_i - 1)}
  & \mathrm{for}\, \, \,  
N \ll 8 \left( \frac{m (\gamma_i - 1)}{g B l_{\ro{c}}} \right)^2 ,
 \\
\sqrt{\frac{N}{2}} \, g B l_{\ro{c}}
  & \mathrm{for}\, \, \,  
N \gg 8 \left( \frac{m (\gamma_i - 1)}{g B l_{\ro{c}}} \right)^2 ,
 \end{dcases}
\label{eq:maru-a}
\end{equation}
where $\gamma_i$ is the initial Lorentz factor of the monopole upon entering the first cell.\footnote{It would be very interesting to study more realistic models where the directions of magnetic fields are not completely random; this should realize a more efficient acceleration of monopoles and thus yield stronger flux bounds.
The analysis here also neglects the effect of the galaxy's gravitational potential, as well as the possibility that the monopoles spend ample time in galactic regions without magnetic fields.} 
In the first line the energy gain is smaller than the initial kinetic energy, i.e. $\Delta E_N < m (\gamma_i -1)$,
while in the second line the monopole has been sufficiently accelerated such that $\Delta E_N > m (\gamma_i -1)$.
If $m (\gamma_i - 1) \ll g B l_{\ro{c}}$,
the energy gain is given by the second line from the first cell.
For the first line of (\ref{eq:maru-a}) to describe well the average behavior of a set of monopoles, the product of the number of monopoles~$p$ and the number of cells~$N$ each monopole passes through need to be large enough such that
\begin{equation}
 p N \gg 16 \left( \dfrac{m (\gamma_i - 1)}{g B l_{\ro{c}}} \right)^2 .
\label{eq:pN}
\end{equation}
The second line of (\ref{eq:maru-a}) works for $p \gg 1$.

\subsection{Do monopoles cluster with a galaxy?}
\label{sec:bound}

If monopoles are bound in a galaxy, they would be moving with the virial velocity~$v_{\ro{vir}}$~($\ll 1$). However since the monopoles, on average, are constantly accelerated in galactic magnetic fields, they will eventually acquire a large enough velocity to escape from the galaxy.
Considering that the escape velocity is not much larger than the virial velocity, let us estimate the time scale for the monopoles to escape from the galaxy as the time it takes for the monopoles' velocity to 
become larger than~$v_{\ro{vir}}$ by a factor of order unity.

If $m v_{\ro{vir}}^2 / 2 \ll g B l_{\ro{c}}$, then the monopole is accelerated to the escape velocity within a single cell.
Then it suffices to consider a uniform magnetic field, in which the velocity varies as $\Delta \bd{v} = g \bd{B} \Delta t / m$ while the monopole is nonrelativistic.
Hence we can estimate the escape time as
\begin{equation}
 \tau_{\ro{esc}} \sim \frac{m v_{\ro{vir}}}{g B}.
\label{eq:esc-I}
\end{equation}
On the other hand if $m v_{\ro{vir}}^2 / 2 \gg g B l_{\ro{c}}$, the monopoles pass through multiple cells before reaching the escape velocity.
The two limiting expressions in (\ref{eq:maru-a}) represent the regimes where the monopole velocity has barely/significantly increased from its initial velocity. The escape velocity is acquired in between the two regimes, when the number of cells passed through is
\begin{equation}
 N_{\ro{esc}} \sim 2 \left(
\frac{m v_{\ro{vir}}^2}{g B l_{\ro{c}}}
\right)^2.
\end{equation}
Hence the escape time is
\begin{equation}
 \tau_{\ro{esc}} \sim \frac{N_{\ro{esc}} l_{\ro{c}}}{v_{\ro{vir}}}
\sim
\frac{2 m^2 v_{\ro{vir}}^3}{g^2 B^2 l_{\ro{c}}}.
\label{eq:esc-II}
\end{equation}
The escape time for both cases $m v_{\ro{vir}}^2 / 2 \ll g B l_{\ro{c}}$ and $m v_{\ro{vir}}^2 / 2 \gg g B l_{\ro{c}}$ can collectively be written as
\begin{equation}\label{eq:tau_esc}
\begin{split}
 \tau_{\ro{esc}} \sim 
\ro{max.} \biggl\{
&\frac{m v_{\ro{vir}}}{g B},
\frac{2 m^2 v_{\ro{vir}}^3}{g^2 B^2 l_{\ro{c}}}
\biggr\}
\\
\sim \ro{max.} \Biggl\{
 &10^7\, \ro{yr} 
\left(\frac{m}{10^{17}\, \ro{GeV}} \right)
\left( \frac{g}{g_{\ro{D}}} \right)^{-1 }
\left( \frac{B}{10^{-6}\, \ro{G}} \right)^{-1}
\left( \frac{v_{\ro{vir}}}{10^{-3}} \right),
\\
 &10^7\, \ro{yr} 
\left(\frac{m}{10^{17}\, \ro{GeV}} \right)^2
\left( \frac{g}{g_{\ro{D}}} \right)^{-2 }
\left( \frac{B}{10^{-6}\, \ro{G}} \right)^{-2}
\left( \frac{l_{\ro{c}}}{1 \, \ro{kpc}} \right)^{-1}
\left( \frac{v_{\ro{vir}}}{10^{-3}} \right)^3
\Biggr\}.
\end{split}
\end{equation}

The escape time decreases as $B$ is amplified, given that the other parameters do not change as much as~$B$.
Monopoles can thus stay clustered with a galaxy if the escape time is longer than the time elapsed since the magnetic field achieved its present-day strength~$B_0$, i.e.,\footnote{The derivation of $\tau_{\ro{esc}}$ uses the assumption of a constant~$B$, which breaks down if $\tau_{\ro{esc}} > \tau_{\ro{sat}}$. In such cases the exact value of $\tau_{\ro{esc}}$ can be modified from (\ref{eq:tau_esc}), but we can still conclude that the monopoles can cluster with the galaxy.
A similar discussion applies to the magnetic field dissipation time which we derive later.}
\begin{equation}
 \left. \tau_{\ro{esc}} \right|_{B = B_0}
>  \tau_{\ro{sat}}.
\label{eq:esc-sat}
\end{equation}
Let us assume hereafter that the time scale of dynamo is comparable to or larger than the time it takes for a particle with virial velocity to cross the magnetic field region of the galaxy, 
\begin{equation}
\tau_{\ro{gen}} \gtrsim 
 \frac{R}{v_{\ro{vir}}} \sim 10^7 \, \ro{yr}
\left( \frac{R}{10\, \ro{kpc}} \right)
\left( \frac{v_{\ro{vir}}}{10^{-3}} \right)^{-1}.
\label{eq:maru-u}
\end{equation}
From this it follows that
$\tau_{\ro{sat}} > l_{\ro{c}} / v_{\ro{vir}}$, 
indicating that monopoles that obtain the escape velocity within a single cell cannot stay clustered until today. 
Hence for monopoles to be clustered, $m v_{\ro{vir}}^2 / 2 \gg g B l_{\ro{c}}$ is a necessary condition.
An even stronger condition is obtained by substituting (\ref{eq:esc-II}) into~(\ref{eq:esc-sat}), which yields a lower bound on the mass of clustered monopoles as
\begin{equation}
m \gtrsim 10^{18}\, \ro{GeV} 
\left( \frac{g}{g_{\ro{D}}} \right)
\left( \frac{B_0}{10^{-6}\, \ro{G}} \right)
\left( \frac{l_{\ro{c}}}{1 \, \ro{kpc}} \right)^{1/2}
\left(\frac{\tau_{\ro{sat}}}{10^{10}\, \ro{yr}} \right)^{1/2}
\left( \frac{v_{\ro{vir}}}{10^{-3}} \right)^{-3/2}.
\label{eq:maru-h}
\end{equation}
Considering for instance the Milky Way, for which the typical parameters of the magnetic field and virial velocity are shown on the right-hand side as the reference values~\cite{Widrow:2002ud,Arshakian:2008cx,Beck:2012bc}, monopoles with a Dirac charge can be clustered today only if their mass is larger than~$10^{18}\, \ro{GeV}$.\footnote{A similar bound can be obtained by requiring the gravitational acceleration of a monopole with virial velocity on a circular orbit at the radius of the galaxy ($v_{\ro{vir}}^2 / r_{\ro{g}}$), to be larger than the magnetic acceleration ($ g B_0 / m$). This yields
\begin{equation}
 m \gtrsim 10^{18}\, \ro{GeV}
\left( \frac{g}{g_{\ro{D}}} \right)
\left( \frac{B_0}{10^{-6}\, \ro{G}} \right)
\left(\frac{r_{\ro{g}}}{10\, \ro{kpc}} \right)
\left( \frac{v_{\ro{vir}}}{10^{-3}} \right)^{-2}.
\end{equation}}
Producing such ultraheavy monopoles in the postinflation universe presents a challenge for monopoles with charge $g \geq g_D$ to serve as dark matter.
This is no longer the case for minicharged ($g \ll g_D$) monopoles, which can cluster with smaller masses.

\subsection{Backreaction from monopoles}

We now derive bounds on the flux of monopoles inside galaxies by studying the backreaction from the monopoles on galactic magnetic fields.

\subsubsection{Unclustered monopoles}

We start by considering monopoles that are not trapped inside a galaxy but pass through it.
The incident flux of such unclustered monopoles on a galaxy is equivalent to the flux inside the galaxy, from monopole number conservation.\footnote{The velocity and number density upon entering the galaxy can each be different from those inside the galaxy, however their product remains constant. Here we do not consider initially unclustered monopoles becoming clustered, or vice versa. We also neglect monopole-antimonopole annihilation.}
Writing the flux per area per solid angle per time as~$F$, and modeling the magnetic field region of the galaxy by a sphere with radius~$R$, then the number of monopoles passing through the magnetic region per time is $4 \pi^2 R^2 F $. 
(The extra power of~$\pi$ is from integrating over the solid angle on one side of the surface of the magnetic region.)
Each monopole crosses roughly $N = R / l_{\ro{c}}$ cells as it traverses the magnetic region, and on average gains energy of $\Delta E_{N = R / l_{\ro{c}}}$. In turn, the magnetic field loses energy at a rate,
\begin{equation}
 \dot{E}_B \sim
- 4 \pi^2 R^2 F \, \Delta E_{N = R / l_{\ro{c}}}.
\end{equation}
Comparing this with the total magnetic field energy, 
$E_B = (4 \pi R^3 / 3) (B^2 / 2)$,
the time scale for the magnetic field to be dissipated is computed as
\begin{equation}
 \tau_{\ro{dis}} = \frac{E_B}{\abs{\dot{E}_B}} \sim
\ro{max.}
\left\{
\frac{2 m (\gamma_i - 1)}{3 \pi g^2 F l_{\ro{c}} }
, \, 
\frac{B}{3 \sqrt{2} \pi g F } \sqrt{\frac{R}{l_{\ro{c}}}}
\right\},
\label{eq:dis_U}
\end{equation}
where we substituted (\ref{eq:maru-a}) into $\Delta E_{N = R / l_{\ro{c}}}$.
Here $\gamma_i$ is understood as the Lorentz factor of the monopoles with respect to the galaxy, upon galaxy entry.

The backreaction from the monopoles has little effect on the magnetic field evolution if the field amplification by dynamo proceeds at a faster rate,
\begin{equation}
 \tau_{\ro{dis}} > \tau_{\ro{gen}}.
\label{eq:2.13}
\end{equation}
This condition should hold throughout the galactic history for negligible backreaction,\footnote{This guarantees negligible backreaction even after the dynamo saturates, if $\tau_{\ro{gen}}$ also sets the time scale for the magnetic field's deviations from the saturation value to decay. However since the field amplification lives on a finite supply of energy of the galaxy, one may instead require $\tau_{\ro{dis}} > \tau_{\ro{sat}}$, giving a stronger bound.} 
and it translates into an upper bound on the monopole flux,
\begin{equation}\label{eq:maru-j}
\begin{split}
F \lesssim \ro{max.} \Biggl\{
 &10^{-16}\, \ro{cm}^{-2} \ro{sec}^{-1} \ro{sr}^{-1}
\left(\frac{m}{10^{17}\, \ro{GeV}} \right)
\left( \frac{g}{g_{\ro{D}}} \right)^{-2}
\left( \frac{l_{\ro{c}}}{1 \, \ro{kpc}} \right)^{-1}
\left(\frac{\tau_{\ro{gen}}}{10^{8}\, \ro{yr}} \right)^{-1}
\left( \frac{ \gamma_i - 1 }{10^{-6}} \right),
\\
 &10^{-16}\, \ro{cm}^{-2} \ro{sec}^{-1} \ro{sr}^{-1}
\left( \frac{g}{g_{\ro{D}}} \right)^{-1}
\left( \frac{B}{10^{-6}\, \ro{G}} \right)
\left( \frac{R}{l_{\ro{c}}} \right)^{1/2}
\left(\frac{\tau_{\ro{gen}}}{10^{8}\, \ro{yr}} \right)^{-1}
\Biggr\}.
\end{split}
\end{equation}
The first (second) line sets the bound when $m$ is larger (smaller) than the threshold value,
\begin{equation}
\hat{m} \sim 10^{17}\, \ro{GeV}
\left( \frac{g}{g_{\ro{D}}} \right)
\left( \frac{B}{10^{-6}\, \ro{G}} \right)
\left( \frac{l_{\ro{c}}}{1 \, \ro{kpc}} \right)
\left(\frac{R}{l_{\ro{c}}} \right)^{1/2}
\left( \frac{ \gamma_i - 1 }{10^{-6}} \right)^{-1}.
\label{eq:m_j}
\end{equation}
Monopoles with masses smaller than this exit the galaxy with a velocity much larger than their incident velocity~$v_i$.

By using the expression (\ref{eq:maru-a}) for $\Delta E_N$ in the above derivation, it was implicitly assumed that the monopoles each pass through at least one cell within the dissipation time~$\tau_{\ro{dis}}$.
Moreover for small-mass monopoles which gain energy as $\Delta E_N \propto \sqrt{N}$ (cf. second line of (\ref{eq:maru-a})), 
we assumed that the time it takes for the monopoles to cross the entire magnetic region is shorter than~$\tau_{\ro{dis}}$.
These two assumptions are automatically satisfied when the condition (\ref{eq:2.13}) holds along with 
\begin{equation}
\tau_{\ro{gen}} \gtrsim 
 \frac{R}{v_{i}} \sim 10^7 \, \ro{yr}
\left( \frac{R}{10\, \ro{kpc}} \right)
\left( \frac{v_{i}}{10^{-3}} \right)^{-1}.
\label{eq:maru-u_i}
\end{equation}
In other words, the flux bound~(\ref{eq:maru-j}) applies without modification under (\ref{eq:maru-u_i}).

It should also be noted that for the first line of (\ref{eq:maru-a}) to well describe the mean behavior of monopoles, the monopole number needs to be large enough to satisfy~(\ref{eq:pN}). 
The total number of unclustered monopoles passing through the magnetic region before the field is dissipated is
$p = 4 \pi^2 R^2 F \tau_{\ro{dis}}$.
Using also the first term in the far right-hand side of (\ref{eq:dis_U}) for~$\tau_{\ro{dis}}$, and $N = R / l_{\ro{c}}$ for the number of cells 
each monopole crosses, 
then (\ref{eq:pN}) yields an upper bound on the monopole mass,
\begin{equation}
m \lesssim  10^{63}\, \ro{GeV}
\left(\frac{B}{10^{-6}\, \ro{G}} \right)^2
\left( \frac{R}{10 \, \ro{kpc}} \right)^{3}
\left( \frac{ \gamma_i - 1 }{10^{-6}} \right)^{-1}.
\label{eq:maru-n}
\end{equation}
The condition (\ref{eq:pN}) is not necessary when $\tau_{\ro{dis}}$ is given by the second term in (\ref{eq:dis_U}), however even in this case the monopole number $p \sim B R^{5/2} / g l_{\ro{c}}^{1/2}$ should be larger than unity for the derivation of the flux bound to be valid. This requires
\begin{equation}
 B \gtrsim 10^{-52}\, \ro{G}
\left( \frac{g}{g_{\ro{D}}} \right)
\left( \frac{l_{\ro{c}}}{1 \, \ro{kpc}} \right)^{1/2}
\left( \frac{R}{10 \, \ro{kpc}} \right)^{-5/2}.
\label{eq:maru-z}
\end{equation}
This condition is equivalent to requiring that $\hat{m}$ given in (\ref{eq:m_j}) is smaller than the upper mass limit of~(\ref{eq:maru-n}).
The conditions (\ref{eq:maru-n}) and (\ref{eq:maru-z}) seem rather weak, however they can become important when considering systems with extremely weak~$B$, or when constraining extremely massive monopoles such as magnetic black holes.

The magnetic field energy taken away by the monopoles can, in principle, later be returned to the field. Then $\tau_{\ro{dis}}$ would only correspond to the half-period of the energy oscillation between the magnetic field and monopoles, and the flux bound would be invalidated.
However it was pointed out in~\cite{Turner:1982ag,Parker:1987} that for monopoles with charge of $g \sim g_{\ro{D}}$, the galactic magnetic fields cannot be maintained in this way since the oscillations are subject to Landau damping, and also because the oscillations would give features of the field that do not match with observations.
It would be important to analyze whether Landau damping is effective with minicharges, $g \ll g_{\ro{D}}$. We leave this for future work.
We also note that for unclustered monopoles, they may fly away from the galaxy before returning the energy to the field.

\subsubsection{Clustered monopoles}

Monopoles that are bound in a galaxy move with the virial velocity~$v_{\ro{vir}}$, and hence each monopole crosses approximately
$N = v_{\ro{vir}} / l_{\ro{c}}$ cells per unit time.
The energy the monopoles steal from the magnetic field per time per volume is thus
\begin{equation}
 \dot{\rho}_B \sim - n \, \Delta E_{N = v_{\ro{vir}} / l_{\ro{c}}},
\end{equation}
where $n$ is the number density of clustered monopoles.
While a monopole is clustered, its energy follows $\Delta E_N \propto N$ as shown in the first line of~(\ref{eq:maru-a}), with 
$\gamma_i - 1 \simeq v_{\ro{vir}}^2 / 2$.
Taking the ratio with the magnetic energy density $\rho_B = B^2 / 2$, 
and noting that the flux is written as $F = n v_{\ro{vir}} / 4 \pi$, the dissipation time scale is obtained as
\begin{equation}
 \tau_{\ro{dis}} = \frac{\rho_B}{\abs{\dot{\rho}_B}}
\sim \frac{m v_{\ro{vir}}^2 }{4 \pi g^2 F l_{\ro{c}} }.
\label{eq:dis_C}
\end{equation}
This matches up to an order-unity factor with the first expression in (\ref{eq:dis_U}) for unclustered monopoles,\footnote{This is because in both (\ref{eq:dis_C}) and the first expression of (\ref{eq:dis_U}), monopoles gain energy as $\Delta E_N \propto N$, and the number of cells crossed per unit time by all the monopoles in the magnetic region is $ \sim 4 \pi^2 F R^3 / l_{\ro{c}}$.} 
after the replacement $v_i \to v_{\ro{vir}}$.
Hence the requirement of negligible backreaction on the magnetic field, $\tau_{\ro{dis}} > \tau_{\ro{gen}}$, 
yields a flux bound that is similar to the first line of (\ref{eq:maru-j}), but with $v_{\ro{vir}}^2/2$ instead of $\gamma_i - 1$.

The derivation assumes that the monopoles pass through at least one cell before their backreaction becomes relevant,
i.e. $\tau_{\ro{dis}} > l_{\ro{c}} / v_{\ro{vir}}$.
This is automatically satisfied under $\tau_{\ro{dis}} > \tau_{\ro{gen}}$ and the condition~(\ref{eq:maru-u}).
From (\ref{eq:maru-u}) it also follows that
the lower mass limit (\ref{eq:maru-h}) for clustered monopoles
is larger than the threshold mass (\ref{eq:m_j}) where the flux bound for unclustered monopoles switches its behavior, if $v_i = v_{\ro{vir}}$.
The flux bound also requires a monopole number large enough to satisfy (\ref{eq:pN}), which yields a mass limit similar to~(\ref{eq:maru-n}).

Here we ignored the possibility of the monopoles escaping from the galaxy before dissipating the magnetic field,
while in Section~\ref{sec:bound} we ignored the monopoles' backreaction on the magnetic field. By combining the discussions, however, we can say that clustered monopoles need to satisfy both the flux bound and the mass bound~(\ref{eq:maru-h}).
Otherwise, either the galactic magnetic field is dissipated, the monopoles are ejected from the galaxy, or both.\footnote{We may guess what happens by comparing the energy required to eject all monopoles from the galaxy per volume, $\rho_{\ro{ej}} \sim n m v_{\ro{vir}}^2 / 2$, and the magnetic energy density today, $\rho_B = B_0^2 / 2$. The former is larger if $m F \gtrsim B_0^2 / 4 \pi v_{\ro{vir}}$. This threshold matches with the value of $m F$ where the mass lower limit (\ref{eq:maru-h}) and flux upper limit (\ref{eq:maru-j}) becomes equal, up to a factor of $\sim \tau_{\ro{sat}} / \tau_{\ro{gen}}$.}

\subsection{Monopole energy loss in interstellar medium and radiative emission}

In our analysis, we neglected the monopole's interaction with the interstellar medium and radiative energy loss.
Here we provide a rough estimate of the effects and show that they are negligible.

Considering for simplicity a monopole moving more or less along a magnetic field line, its energy gain from the magnetic field per unit length is
\begin{equation}
    \left( \frac{d E}{d x} \right)_{\mathrm{mag}} \sim g B .
\end{equation}
For instance for a charge $g = g_{\mathrm{D}}$ and field strength $B = 10^{-6} \, \mathrm{G}$, the energy gain is $\left(dE / dx \right)_{\mathrm{mag}} \sim 10^{-2}\, \mathrm{eV\, cm^{-1}}$.

Monopoles lose energy in the interstellar medium by ionization and atomic excitation of the constituent neutral particles.
The energy loss can be evaluated from the Bethe--Bloch formula considering monopoles as particles with a velocity-dependent electric charge, $q = g v$ \cite{Ahlen:1980xr}. This gives, at the order-of-magnitude level, 
\begin{equation}
\label{eq:betheBloch}
    - \left( \frac{d E}{d x} \right)_{\mathrm{ion}} \sim \frac{e^2 g^2 n_{\mathrm{m}}}{m_{\mathrm{\mathrm{e}}}} ,
\end{equation}
where $n_{\mathrm{m}}$ is the number density of atoms in interstellar space and $m_{\mathrm{\mathrm{e}}}$ is the electron mass.
Assuming $n_{\mathrm{m}} = 1~\mathrm{cm}^{-3}$ \cite{Longair:2011} and $g = g_{\mathrm{D}}$, the energy loss is $- \left(dE / dx \right)_{\mathrm{ion}} \sim 10^{-14}\, \mathrm{eV\, cm^{-1}}$, which is completely negligible with respect to the energy gain from the magnetic field.
Other electromagnetic processes that induce energy loss of monopoles in matter include pair production and photonuclear interactions, however both contributions are subdominant compared to the ionization effect for $\gamma < 10^{4}$ \cite{Wick:2000yc}.
There can also be energy loss through bremsstrahlung radiation in collisions; however this effect is inversely proportional to the monopole mass~\cite{Wick:2000yc}, so we expect it also to be subdominant.

Monopoles lose energy also by emitting radiation as they are accelerated by the galactic magnetic field.
The energy loss by this process can be described by the magnetic dual of the Larmor formula, which for a nonrelativistic monopole accelerated as $\dot{v} \sim g B / m$ yields,
\begin{equation}
\label{eq:larmor}
    - \left( \frac{d E}{d t} \right)_{\mathrm{rad}} \sim \frac{g^4 B^2}{m^2}.
\end{equation}
Assuming  $B = 10^{-6}~\mathrm{G}$, $g = g_{\mathrm{D}}$, $m = 100~\mathrm{GeV}$, and a monopole velocity $v = 10^{-3}$, the energy loss per unit length is $- \left(dE / dx \right)_{\mathrm{rad}} \sim 10^{-25} \, \mathrm{eV \, cm^{-1}}$. Here we considered the smallest monopole mass admitted by the bound from \cite{MoEDAL:2021vix} to maximize the energy loss, however it is still negligible compared to the energy gain from the magnetic field.

We expect the main results of this section to be generic, however it would be important to analyze energy losses in more realistic models of the distribution of the interstellar medium, and also to perform a systematic study in the full parameter space. 
We leave these for the future.

\subsection{Summary of bounds from galactic magnetic fields}
\label{sec:summGal}

We have seen that the bounds on the flux of clustered and unclustered monopoles inside galaxies are collectively described by (\ref{eq:maru-j}),
given that the dynamo time scale, monopole mass, and magnetic field respectively satisfy (\ref{eq:maru-u_i}), (\ref{eq:maru-n}), and (\ref{eq:maru-z}).
For unclustered monopoles $\gamma_i$ in these expressions denotes
the initial Lorentz factor with respect to the galaxy, 
while for clustered monopoles it is given by the virial velocity as
$\gamma_i - 1 = v_{\ro{vir}}^2 / 2$.
Clustered monopoles further need to satisfy the lower bound on the mass~(\ref{eq:maru-h}) in order to stay clustered until today.

The flux bound (\ref{eq:maru-j}) at large~$m$ increases with~$m$ whereas it is independent of~$B$, and vice versa at small~$m$.
Considering present-day magnetic fields, 
whose amplitude in spiral galaxies is typically of $B_0 \sim 10^{-6}\, \mathrm{G}$, one reproduces the results of~\cite{Turner:1982ag} (see also \cite{Bai:2020spd,Graesser:2021vkr}).
However, the bound applies throughout the history of a galaxy, and thus the bound at low masses can be improved by studying galaxies in the past when their magnetic fields were weaker.
Strong bounds are obtained from the initial seed field for galactic dynamo~\cite{Adams:1993fj},\footnote{The results in \cite{Turner:1982ag} and \cite{Adams:1993fj} are slightly different at the high mass end where the bound is independent of~$B$; this is because the two works use different values for the other parameters such as~$l_{\ro{c}}$, and also different rounding methods.} 
although there is a huge uncertainty in the seed field ranging typically between $10^{-30}\, \ro{G} \lesssim B \lesssim 10^{-10}\, \ro{G}$~\cite{Widrow:2002ud,Arshakian:2008cx,Beck:2012bc}.
We also note that increasing~$l_{\ro{c}}$ and/or $g$ improves the flux bound, as well as the lower mass limit for clustered monopoles.

\begin{figure}[!t]
  \centering
  \includegraphics[width=0.9\textwidth]{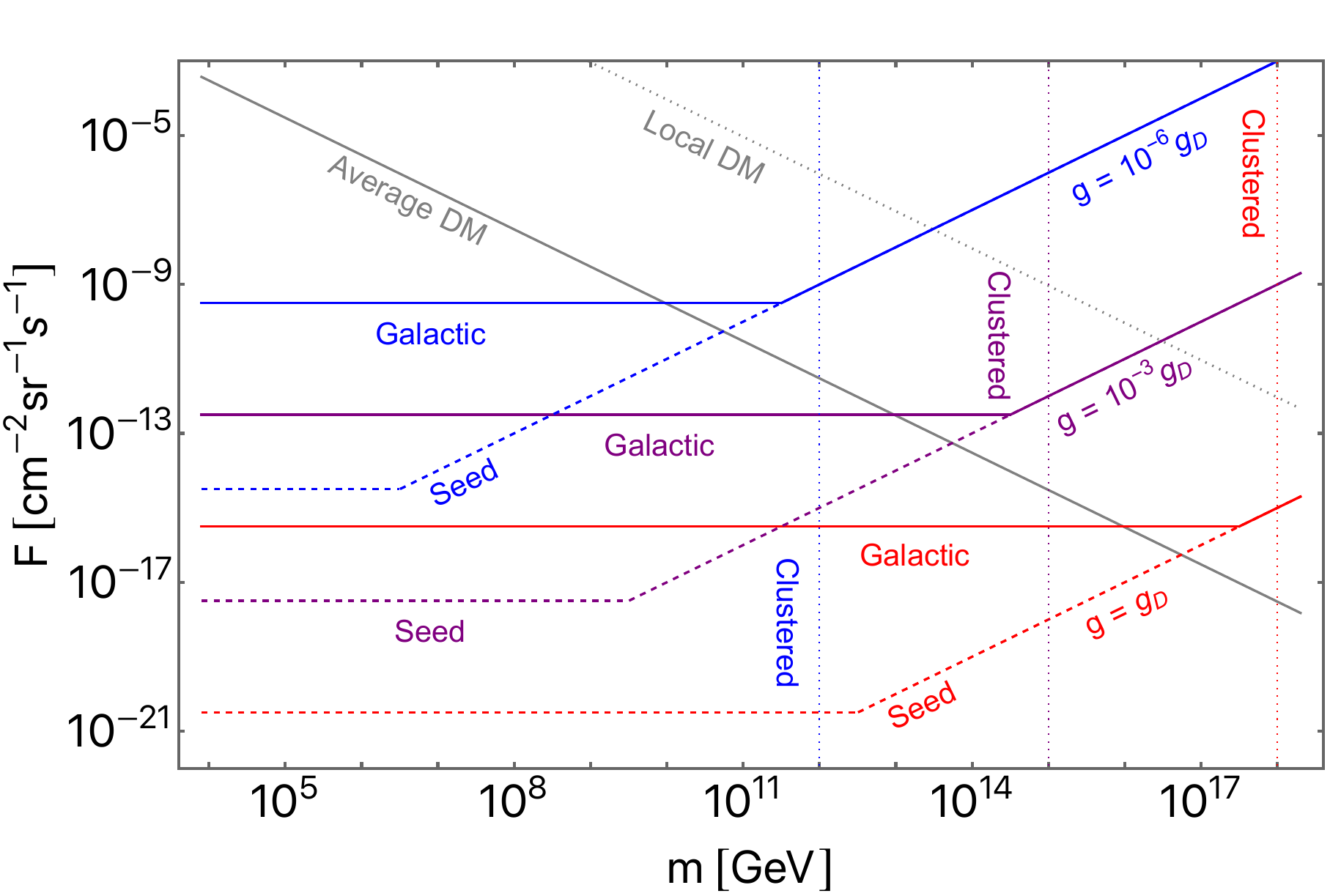}
\caption{
Upper bound \eqref{eq:maru-j} on the monopole flux as a function of mass, from the survival of Galactic magnetic fields ($B = 10^{-6}\, \ro{G}$, solid lines) and seed fields ($B = 10^{-11}\, \ro{G}$, dashed). 
The magnetic charge is varied as $g = g_{\mathrm{D}}$ (red), $10^{-3} g_{\mathrm{D}}$ (purple), $10^{-6} g_{\mathrm{D}}$ (blue). 
The lower mass limit (\ref{eq:maru-h}) for monopoles to stay clustered with the Galaxy is shown by the vertical dotted lines with different colors corresponding to different~$g$.
Other parameters are fixed to $l_{\mathrm{c}} = 1\, \mathrm{kpc}$, $R = 10\, \mathrm{kpc}$, $\tau_{\mathrm{gen}} = 10^8\, \mathrm{yr}$, 
$\tau_{\mathrm{sat}} = 10^{10}\, \mathrm{yr}$, and $\gamma_i - 1 = 10^{-6}$.
Also shown are bounds from requiring the density of monopoles not to exceed that of dark matter, for unclustered (gray solid) and clustered (gray dotted) monopoles.
}
\label{fig:Parker}
\end{figure}

In Figure~\ref{fig:Parker} we show the flux upper bound \eqref{eq:maru-j} as a function of the monopole mass, with the magnetic charge varied as $g = g_{\mathrm{D}}$ (red), $10^{-3} g_{\mathrm{D}}$ (purple), $10^{-6} g_{\mathrm{D}}$ (blue). 
The solid lines denote bounds from the magnetic field in the present Milky Way, taken as $B = 10^{-6}\, \ro{G}$. The dashed lines show how the bound improves by considering a seed field of $B = 10^{-11}\, \ro{G}$.
The dotted vertical lines represent the lower mass limit (\ref{eq:maru-h}) of clustered monopoles in the Milky Way.
Here the other parameters are taken as $l_{\mathrm{c}} = 1\, \mathrm{kpc}$, $R = 10\, \mathrm{kpc}$, $\tau_{\mathrm{gen}} = 10^8\, \mathrm{yr}$, 
$\tau_{\mathrm{sat}} = 10^{10}\, \mathrm{yr}$, and $\gamma_i - 1 = 10^{-6}$.

In the plot we also show bounds from the requirement that the density of monopoles~$\rho_{\ro{M}}$ does not exceed the dark matter density~$\rho_{\ro{DM}}$. 
Using $\rho_{\ro{M}} = m n$ for nonrelativistic monopoles 
with $n$ being the number density, 
the requirement translates into an upper bound on the monopole flux $F = n v_i / 4 \pi$ as,
\begin{equation}\label{eq:cosmological}
\begin{split}
 F & \leq \frac{\rho_{\ro{DM}} v_i }{4 \pi m} \\
 & \approx 3 \times 10^{-17}\, \ro{cm}^{-2} \ro{sec}^{-1} \ro{sr}^{-1} \left(\frac{m}{10^{17}\, \ro{GeV}}\right)^{-1} \left( \frac{v_i}{10^{-3}} \right)
\left( \frac{\rho_{\ro{DM} } }{1.3 \times 10^{-6} \, \ro{GeV} \, \ro{cm}^{-3}}  \right).
\end{split}
\end{equation}
The flux of unclustered monopoles is bound by setting the dark matter density to the average value in the universe,
$\rho_{\ro{DM}} \approx 1.3 \times 10^{-6} \, \ro{GeV} \, \ro{cm}^{-3}$~\cite{Planck:2018vyg}; this is shown in the plot as the gray solid line. On the other hand, the abundance of clustered monopoles should be compared to the local dark matter density in galaxies;
the gray dotted line shows the bound using the value in our Milky Way,
$\rho_{\ro{DM}} \approx 0.4 \,\ro{GeV} \, \ro{cm}^{-3}$~\cite{Read:2014qva}.

One sees in the plot that for clustered monopoles, the bound from the local dark matter density (which scales as $\propto m^{-1}$) is stronger than that from the survival of galactic fields ($\propto m$) for most of the mass range where the monopoles can be clustered. This can be shown explicitly by comparing the mass~$m_{\ro{eq}}$ where the two upper bounds ((\ref{eq:cosmological}) and the first line of (\ref{eq:maru-j})) become equal, to the lower limit on the mass~$m_{\ro{cl}}$ for clustered monopoles (cf.~(\ref{eq:maru-h})); their ratio is
\begin{equation}
\frac{m_{\ro{eq}}}{m_{\ro{cl}}}
\sim 10
\left( \frac{B_0}{10^{-6}\, \ro{G}} \right)^{-1}
\left(\frac{\tau_{\ro{gen}}}{10^{8}\, \ro{yr}} \right)^{1/2}
\left(\frac{\tau_{\ro{sat}}}{10^{10}\, \ro{yr}} \right)^{-1/2}
\left( \frac{v_{\ro{vir}}}{10^{-3}} \right)
\left( \frac{\rho_{\ro{DM} } }{0.4 \, \ro{GeV} \, \ro{cm}^{-3}}  \right)^{1/2}.
\end{equation}
This shows that $m_{\ro{eq}}$ and $m_{\ro{cl}}$ are not too different
in the Milky Way whose magnetic field and dark matter parameters are typically given by the reference values in the right-hand side. This means that if monopoles can cluster with our Galaxy and their density does not exceed that of dark matter, then they almost automatically satisfy the Parker bound from Galactic fields.\footnote{It would be interesting to understand whether $m_{\ro{eq}} \sim m_{\ro{cl}}$ holds for generic galaxies hosting magnetic fields, by studying the relation between the dark matter density and the dynamo action.}
In the literature the Galactic Parker bound has often been analyzed for constraining monopoles as a dark matter candidate; however most such studies focus on parameter regions where the monopoles actually cannot cluster with our Galaxy and hence obviously cannot serve as dark matter.

\section{Bounds from primordial magnetic fields}
\label{sec:PrimParker}

In this section we extend the computations for the bounds from primordial magnetic fields derived in \cite{Long:2015cza, Kobayashi:2022qpl} to allow for the monopoles to carry arbitrary magnetic charge. 

Magnetic monopoles are accelerated by the primordial magnetic fields and the fields consequently lose their energy. If the interaction between the monopoles and the charged particles of the primordial plasma is sufficiently strong, the energy of the primordial magnetic fields is eventually transferred to the primordial plasma. From the requirement that the primordial magnetic fields survive until today, we get a bound on the abundance of monopoles. 
On the other hand, if the interaction between the monopoles and the primordial plasma is weak, then the energy oscillates between the monopoles and the magnetic fields.
This modifies the time evolution of the magnetic fields \cite{Long:2015cza}, however it does not lead to a dissipation of the fields.
Thus a bound on monopoles is obtained if their interaction with the primordial plasma is sufficiently strong at least for some period in the early universe.

We first describe the evolution of the primordial magnetic fields in the presence of monopoles with arbitrary magnetic charge from the end of magnetogenesis to the epoch of $e^+ e^-$ annihilation, when the number of charged particles in the universe becomes drastically reduced. 
Then, we derive bounds on the monopole abundance from the survival of the primordial magnetic fields.
We consider a Friedmann-Robertson-Walker (FRW) background spacetime: $ds^2 = dt^2 - a^2 d \bd{x}^2$.

In our analysis we suppose that the process of magnetogenesis terminates at the end of inflation or during the reheating phase. Thus, we study the dynamics of the primordial magnetic fields during the reheating epoch when the energy density of the universe is dominated by an oscillating inflaton field, and the subsequent epoch of radiation domination. 
We define $T_{\mathrm{dom}}$ as the temperature at the end of reheating when the universe becomes dominated by radiation (the subscript ``$\mathrm{dom}$'' denotes quantities computed at this time).
We use the subscript ``end'' to denote quantities computed at the end of magnetogenesis.\footnote{Notice that in \cite{Kobayashi:2022qpl} we referred to the time at the end of magnetogenesis as $t_i$, instead of $t_{\mathrm{end}}$.
Moreover, we used $n$ to denote the number density of monopole-antimonopole pairs, while in this paper we will use it for the total number density of monopoles and antimonopoles, i.e. $n \to n/2$.}
In the absence of any source, the energy density of primordial magnetic fields $\rho_{\mathrm{B}}$ redshifts simply as radiation, $\rho_{\mathrm{B}} \propto a^{-4}$. 
The magnetic field amplitude thus redshifts as $B \propto a^{-2}$, since $\rho_{\mathrm{B}} = B^2/2$.

The existence of intergalactic magnetic fields with strength $B_0 \gtrsim 10^{-15}~\mathrm{G}$ (we use the subscript ``$0$'' to denotes quantities in the present universe), and coherence length of $\mathrm{Mpc}$ scale or larger has been suggested by gamma ray observations \cite{Tavecchio:2010mk, Neronov:2010gir, Dermer:2010mm}.
Such scales have always been outside the Hubble horizon during the period from the end of inflation to $e^+ e^-$ annihilation.
Thus, since the distance crossed by the monopoles during the period of interest is smaller than the correlation length of the fields, we treat the fields as effectively homogeneous.
In this paper we use $10^{-15}~\mathrm{G} \simeq 2 \cdot 10^{-17}~\mathrm{eV^2}$ as the reference value for the intergalactic magnetic field strength today.

Under the above assumptions, the evolution of the energy density of primordial magnetic fields in the presence of monopoles is described by the equation:
\begin{equation}
\label{piEq}
    \frac{\dot{\rho}_{\mathrm{B}}}{\rho_{\mathrm{B}}}=-\Pi_{\mathrm{red}} - \Pi_{\mathrm{acc}} ,
\end{equation}
where an overdot denotes a derivative with respect to physical time~$t$. Here $\Pi_{\mathrm{red}}$ and $\Pi_{\mathrm{acc}}$ are the dissipation rates of the magnetic field energy due to redshifting and monopole acceleration:
\begin{subequations} \label{PiAll}
\begin{equation}
\label{PiRed}
\Pi_{\mathrm{red}} = 4 H,
\end{equation}
\begin{equation}
\Pi_{\mathrm{acc}} = \frac{2 g}{B} n v ,
\end{equation}
\end{subequations}
where $H = \dot{a}/a$ is the Hubble rate, $n$ is the physical number density of monopoles, and $v$ is the velocity of the monopoles. We neglect the production of monopole pairs by the magnetic fields through the Schwinger effect \cite{Schwinger:1951nm, Affleck:1981ag, Affleck:1981bma,Kobayashi:2021des,Kobayashi:2022qpl}. Thus, we assume that the comoving number density is constant in time\footnote{See \cite{Kobayashi:2021des} for detailed discussions on monopoles produced by the primordial magnetic field itself.}, i.e. $n \propto a^{-3}$.
The expression for the ratio $\Pi_{\mathrm{acc}} / \Pi_{\mathrm{red}}$ then can be written as:
\begin{equation}
\label{eq:ratio}
    \frac{\Pi_{\mathrm{acc}}}{\Pi_{\mathrm{red}}} = \frac{g}{2 B H} n v .
\end{equation}

We require the condition $\Pi_{\mathrm{acc}} / \Pi_{\mathrm{red}} \ll 1$ to hold during the period from the end of magnetogenesis, $t = t_{\mathrm{end}}$, to $e^+ e^-$ annihilation. This condition corresponds to having negligible backreaction on the primordial magnetic fields from the monopole acceleration. In order to rewrite such a condition as a bound on the monopole abundance, in the next section we study the evolution of the monopole velocity in the early universe.

\subsection{Monopole dynamics in primordial magnetic fields}
\label{sec:3.1}

We now describe the motion of monopoles accelerated by a homogeneous magnetic field in the presence of a primordial plasma.

For the analysis we suppose the plasma to be at rest in the coordinate system $(t, x^i)$. 
We also ignore monopole velocities perpendicular to the direction of the magnetic field because such velocity components decay away. 
Further ignoring random thermal velocities, the motion of monopoles with magnetic charge $g$ and mass $m$ can be described by the equation~\cite{Kobayashi:2022qpl}:
\begin{equation}
\label{eq:eqVelocity}
    m \frac{d}{dt} ( \gamma v) = g B - \left( f_{\mathrm{p}} + m H \gamma \right) v .
\end{equation}
The term $f_{\mathrm{p}}$ is proportional to the cross section of the interaction between the monopoles and the charged particles in the plasma. When the particles in the plasma are relativistic and in thermal equilibrium, $f_{\mathrm{p}}$ can be expressed as \cite{Vilenkin:2000jqa}:
\begin{equation}
\label{effep}
    f_{\mathrm{p}} \sim \frac{e^2 g^2 \mathcal{N}_c}{16 \pi^2} T^2 .
\end{equation}
Here $T$ is the temperature of the plasma and $\mathcal{N}_c$ is the effective number of relativistic and electrically charged degrees of freedom in thermal equilibrium including also the contributions of the spin and the charge of the scatterers.
In Eq.~\eqref{eq:eqVelocity}, the expansion of the universe can be seen as an additional frictional term proportional to the Hubble rate.
In \cite{Kobayashi:2022qpl} the solution of the equation of motion has been studied for magnetic monopoles with Dirac charge $g_{\mathrm{D}} = 2 \pi /e$. Here we are interested in generalizing the results to generic magnetic charges.

Depending on the parameters one of the two frictional terms becomes dominant and eventually the monopoles achieve a terminal velocity.
If the Hubble friction is the dominant term, i.e. $mH \gamma \gg f_{\mathrm{p}}$, the terminal velocity is set approximately by:
\begin{equation}
\label{eq:vHubble}
    \left( \gamma v \right)_{\mathrm{H}} \sim \frac{g B}{m H} .
\end{equation}
The expression is directly proportional to the magnetic charge of the monopoles. Thus, smaller magnetic charge corresponds to smaller $v_{\mathrm{H}}$.

On the other hand, when the drag force by the interaction with the plasma is dominant, i.e. $mH \gamma \ll f_{\mathrm{p}}$, and the monopoles move at nonrelativistic velocities, the terminal velocity corresponds to:
\begin{equation}
\label{eq:vPlasma}
    v_{\mathrm{p}} = \frac{gB}{f_{\ro{p}}}
\sim \frac{16 \pi^2 B}{e^2 g \mathcal{N}_c T^2} .
\end{equation}
Since the interaction rate with the particles of the plasma is proportional to $g^2$ and the monopole acceleration by the magnetic field to $g$, the velocity $v_{\mathrm{p}}$ scales as $v_{\mathrm{p}} \propto g^{-1}$.

Due to the $\gamma$ factor in front of the Hubble friction term, for relativistic monopoles ($\gamma \gg 1$) the drag force due to the expansion of the universe tends to become dominant. 
In this case the terminal velocity of the monopoles corresponds to the value of $v_{\mathrm{H}}$ shown in Eq.~\eqref{eq:vHubble}.
However, in the case when the monopoles move at relativistic velocities and $mH \gamma \ll f_{\mathrm{p}}$, the monopole velocity rapidly decreases to nonrelativistic values and eventually starts to follow the terminal velocity $v_{\mathrm{p}}$ \cite{Kobayashi:2022qpl}.

\begin{figure}[!t]
  \centering
  \begin{subfigure}[b]{0.496\textwidth}
    \centering
    \includegraphics[width=\textwidth]{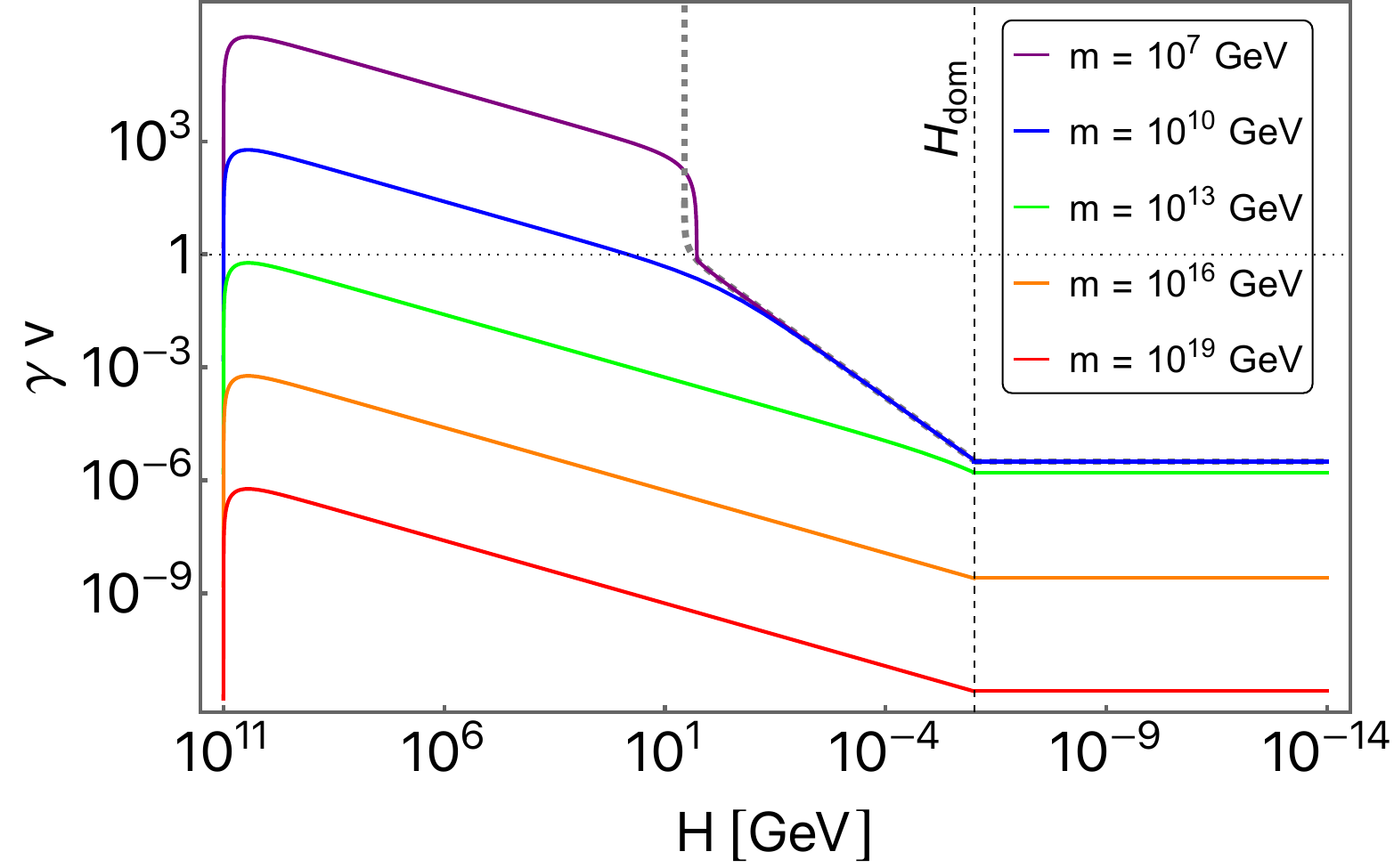}
    \caption{$g = 10^{-3} g_{\mathrm{D}}$ with varying $m$.}
    \label{fig:velocity}
  \end{subfigure}
  \hfill
  \begin{subfigure}[b]{0.496\textwidth}
    \centering
    \includegraphics[width=\textwidth]{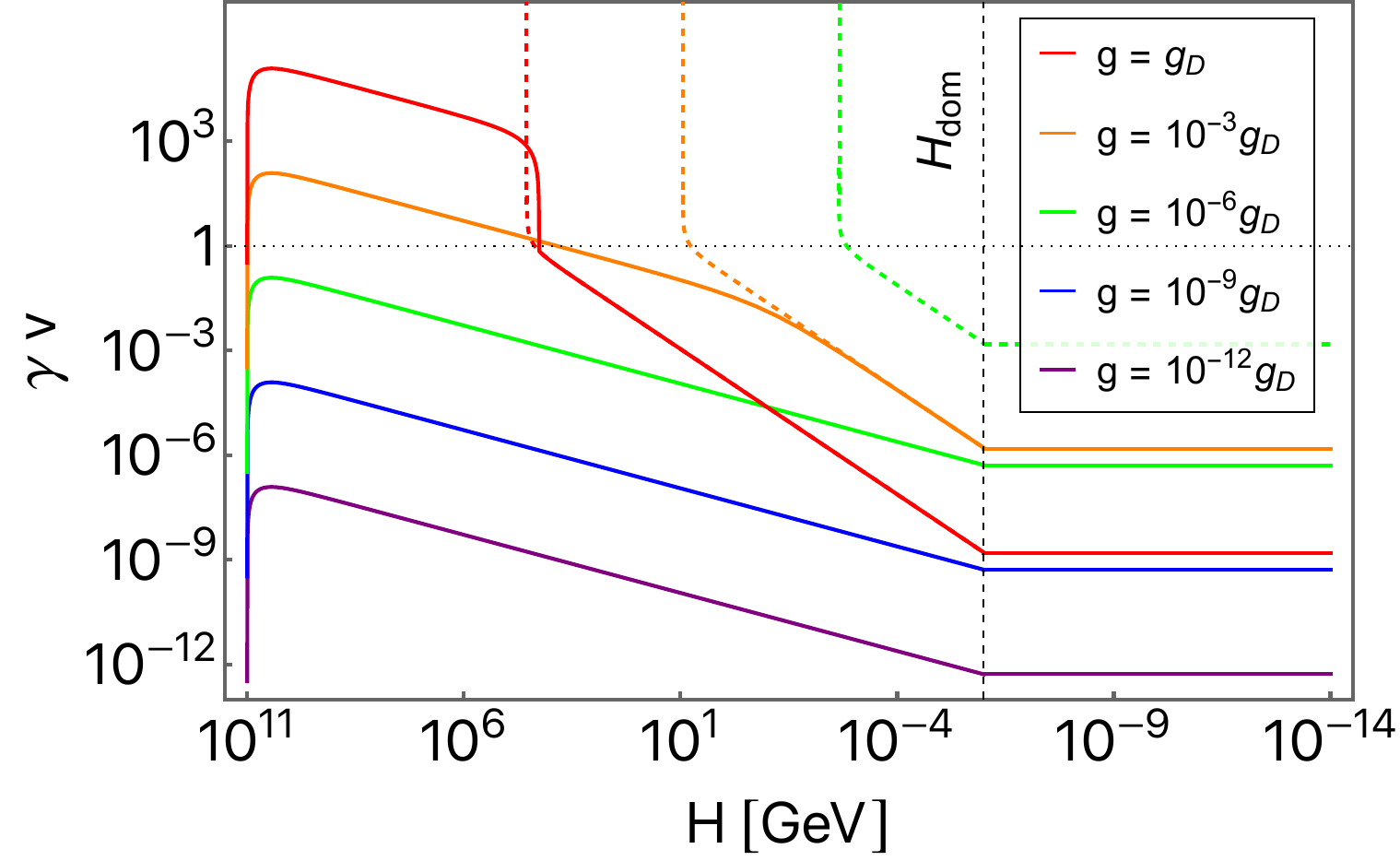}
    \caption{$m = 10^{11}\, \ro{GeV}$ with varying $g$.}
    \label{fig:velocity2}
  \end{subfigure}
  \hfill
  \begin{subfigure}[b]{0.496\textwidth}
    \centering
    \includegraphics[width=\textwidth]{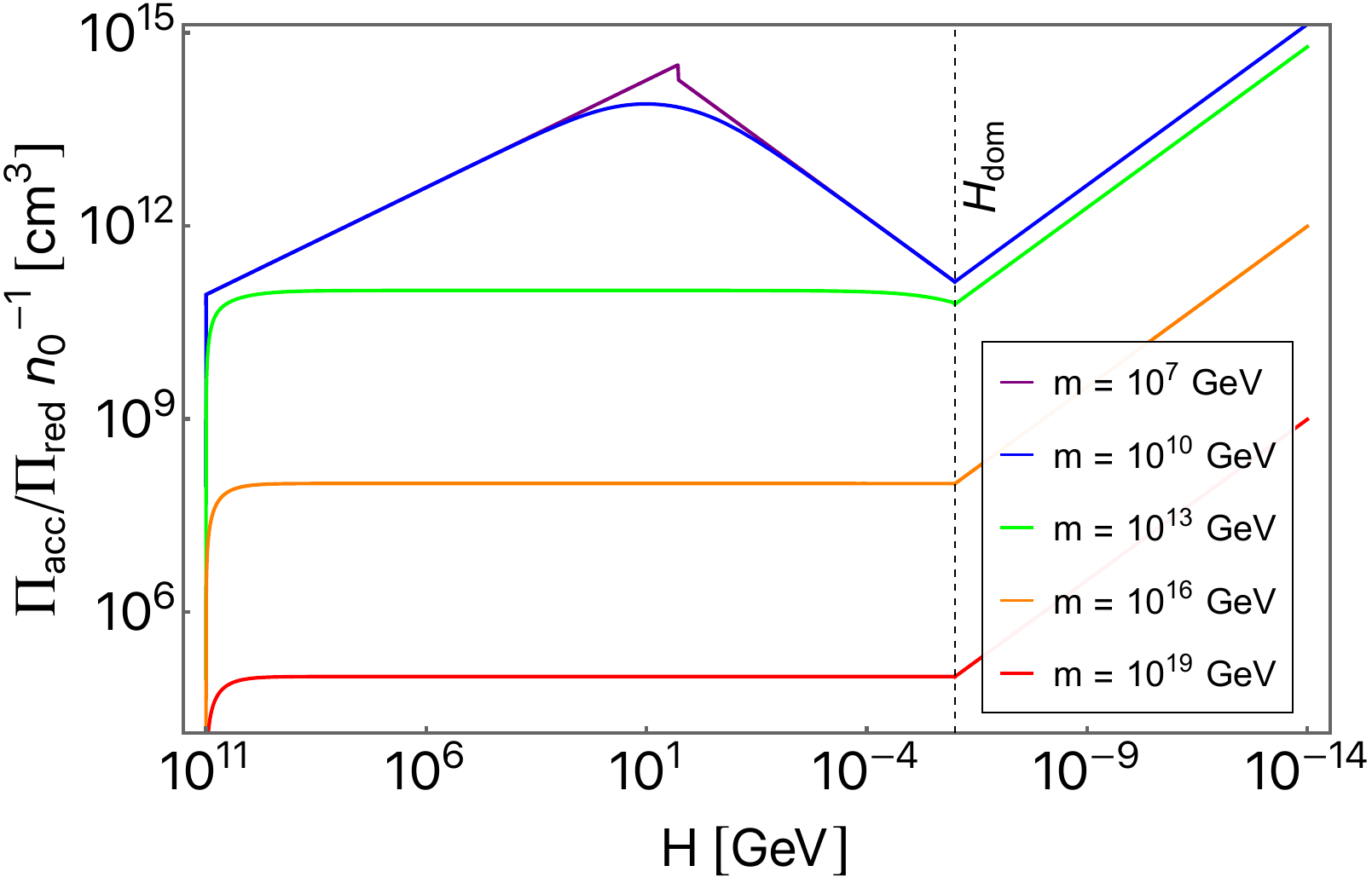}
    \caption{$g = 10^{-3} g_{\mathrm{D}}$ with varying $m$.}
    \label{fig:ratio}
  \end{subfigure}
  \hfill
  \begin{subfigure}[b]{0.496\textwidth}
    \centering
    \includegraphics[width=\textwidth]{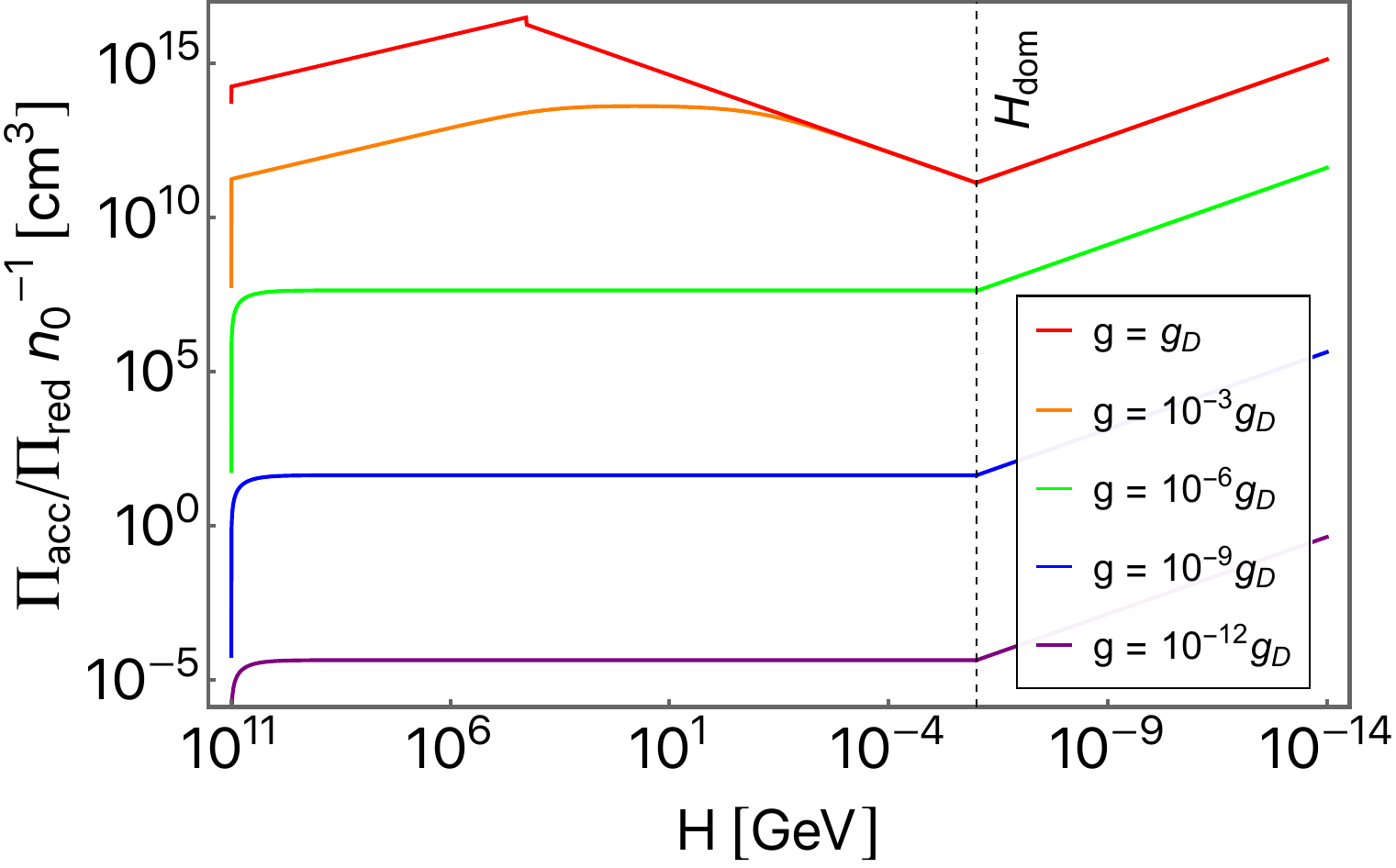}
    \caption{$m = 10^{11}\, \ro{GeV}$ with varying $g$.}
    \label{fig:ratio2}
  \end{subfigure}
  \caption{Time evolution of the monopole velocity in primordial magnetic fields (upper panels) and the normalized dissipation rate of the magnetic fields due to monopole acceleration (lower panels).
The Hubble scales at the end of magnetogenesis and at the onset of radiation domination are taken respectively as
$H_{\mathrm{end}} = 10^{11}\, \mathrm{GeV}$ and $H_{\mathrm{dom}} = 10^{-6}\, \mathrm{GeV}$. 
The present-day magnetic field strength is taken as
$B_0 = 10^{-15}\, \mathrm{G}$. 
The numbers of relativistic (charged) degrees of freedom are fixed to $g_{*} = \mathcal{N}_c = 100$.
In the left panels, the charge of the monopole is fixed to $g = 10^{-3} g_{\mathrm{D}}$ while the mass is varied as 
$m = 10^{19}\, \mathrm{GeV}$ (red), $10^{16}\, \mathrm{GeV}$ (orange), $10^{13}\, \mathrm{GeV}$ (green), $10^{10}\, \mathrm{GeV}$ (blue), $10^{7}\, \mathrm{GeV}$ (purple), from bottom to top. 
In the right panels the mass is fixed to $m = 10^{11}\, \mathrm{GeV}$ while the charge is varied as $g = g_{\mathrm{D}}$ (red), $10^{-3} g_{\mathrm{D}}$ (orange), $10^{-6} g_{\mathrm{D}}$ (green), $10^{-9} g_{\mathrm{D}}$ (blue), $10^{-12} g_{\mathrm{D}}$ (purple), from top to bottom. The dashed colored curves in the upper panels show the terminal velocity set by the friction from the cosmological plasma.
}
  \label{fig:VelRatio}
\end{figure}

In Figure~\ref{fig:VelRatio} we plot the time evolution of $\gamma v$ (Figures~\ref{fig:velocity} and~\ref{fig:velocity2}) and of the ratio $\Pi_{\mathrm{acc}} / \Pi_{\mathrm{red}}$ normalized by the monopole number density today (Figures~\ref{fig:ratio} and~\ref{fig:ratio2}).
The time evolution of $\gamma v$ is obtained by numerically solving the equation of motion Eq.~\eqref{eq:eqVelocity} with an initial condition of $v_{\ro{end}} = 0$. The time evolution of $\Pi_{\mathrm{acc}} / \Pi_{\mathrm{red}}$ is obtained by substituting into Eq.~\eqref{eq:ratio} the numerical solution of Eq.~\eqref{eq:eqVelocity}. 
For the plots we assume $H_{\mathrm{end}} = 10^{11}~\mathrm{GeV}$,  $H_{\mathrm{dom}} = 10^{-6}~\mathrm{GeV}$ (i.e. $T_{\ro{dom}} \sim 10^6\, \ro{GeV}$), and fix the number of relativistic (charged) degrees of freedom as $g_{*} = \mathcal{N}_c = 100$ throughout the displayed epochs.
The magnetic field strength is taken such that it approaches a present-day strength of $B_0 = 10^{-15}\ \mathrm{G}$.

In Figures~\ref{fig:velocity} and~\ref{fig:ratio}, the results are shown for a magnetic charge $g = 10^{-3} g_{\mathrm{D}}$ and for different values of the monopole mass. The value of the magnetic charge has been chosen in order to cover a wide range of possible behaviors of the monopole velocity which we will explain in the following sections. Each value of the mass is associated to a differently colored solid curve; from bottom to top, red: $m = 10^{19}~\mathrm{GeV}$, orange: $m = 10^{16}~\mathrm{GeV}$, green: $m = 10^{13}~\mathrm{GeV}$, blue: $m = 10^{10}~\mathrm{GeV}$, purple: $m = 10^{7}~\mathrm{GeV}$. The purple curve disappears when it is behind the blue curve.
In Figure~\ref{fig:velocity}, the dashed gray line shows $\gamma v$ with $v$ substituted by $v_{\ro{p}}$ given in Eq.~\eqref{eq:vPlasma}. This corresponds to the terminal velocity set by the plasma when $v_{\mathrm{p}} \ll 1$, and it overlaps with the blue and purple lines in the right part of the figure.

In Figures~\ref{fig:velocity2} and~\ref{fig:ratio2}, the results are shown for a mass $m = 10^{11}~\mathrm{GeV}$ and for different values of the magnetic charge. As in the previous case, the value of the mass has been chosen in order to show the various behaviors of the monopole velocity. Each value of the charge is associated to a differently colored solid curve; from top to bottom, red: $g = g_{\mathrm{D}}$, orange: $g = 10^{-3} g_{\mathrm{D}}$, green: $g = 10^{-6} g_{\mathrm{D}}$, blue: $g = 10^{-9} g_{\mathrm{D}}$, purple: $g = 10^{-12} g_{\mathrm{D}}$. In Figure~\ref{fig:ratio2} the orange curve disappears when it is behind the red curve.
In Figure~\ref{fig:velocity2}, the dashed curves show $(\gamma v)_{\ro{p}}$ for different charges, indicated by the colors. 
The red and orange dashed curves overlap with the corresponding solid curves in the right part of the plot.
The blue and purple dashed curves are not shown because a terminal velocity set by the friction with the plasma cannot be defined in those cases, being $v_{\mathrm{p}} \gg 1$.

In the figures the monopole velocity follows $v_{\mathrm{p}}$ in Eq.~\eqref{eq:vPlasma} shown as the dashed lines, otherwise it follows $v_{\mathrm{H}}$ in Eq.~\eqref{eq:vHubble} (except for at the left edges of the plots where $H \sim H_{\ro{end}}$\footnote{Radiative emission may affect the monopole dynamics before one of the terminal velocities is reached. However this also depends on how the magnetic field is initially switched on.}). 
This indicates that one of the two terminal velocities always gives an attractor solution for the monopole velocity. One also sees from the figures that the velocity can make a transition from $v_{\mathrm{H}}$ to $v_{\mathrm{p}}$ as the universe expands, but not vice versa. The transition can be smooth as for the blue curve in Figure~\ref{fig:velocity}, but can also take the form of a sudden jump as for the purple curve in Figure~\ref{fig:velocity}.

\subsection{Radiation-dominated epoch}
\label{sec:vel:during}

We start by analyzing the backreaction of monopoles on primordial magnetic fields during the radiation-dominated epoch.
Neglecting the time dependence of $g_{*(s)}$ and $\mathcal{N}_c$, then during radiation domination the Hubble rate redshifts as $H \propto a^{-2}$, and the temperature of the plasma as $T \propto a^{-1}$. 
These, together with $B \propto a^{-2}$, render both $v_{\mathrm{p}}$ and $v_{\mathrm{H}}$ constant in time. The monopoles during radiation domination thus move with a constant velocity.

\begin{figure}[!t]
  \centering
  \includegraphics[width=0.7\textwidth]{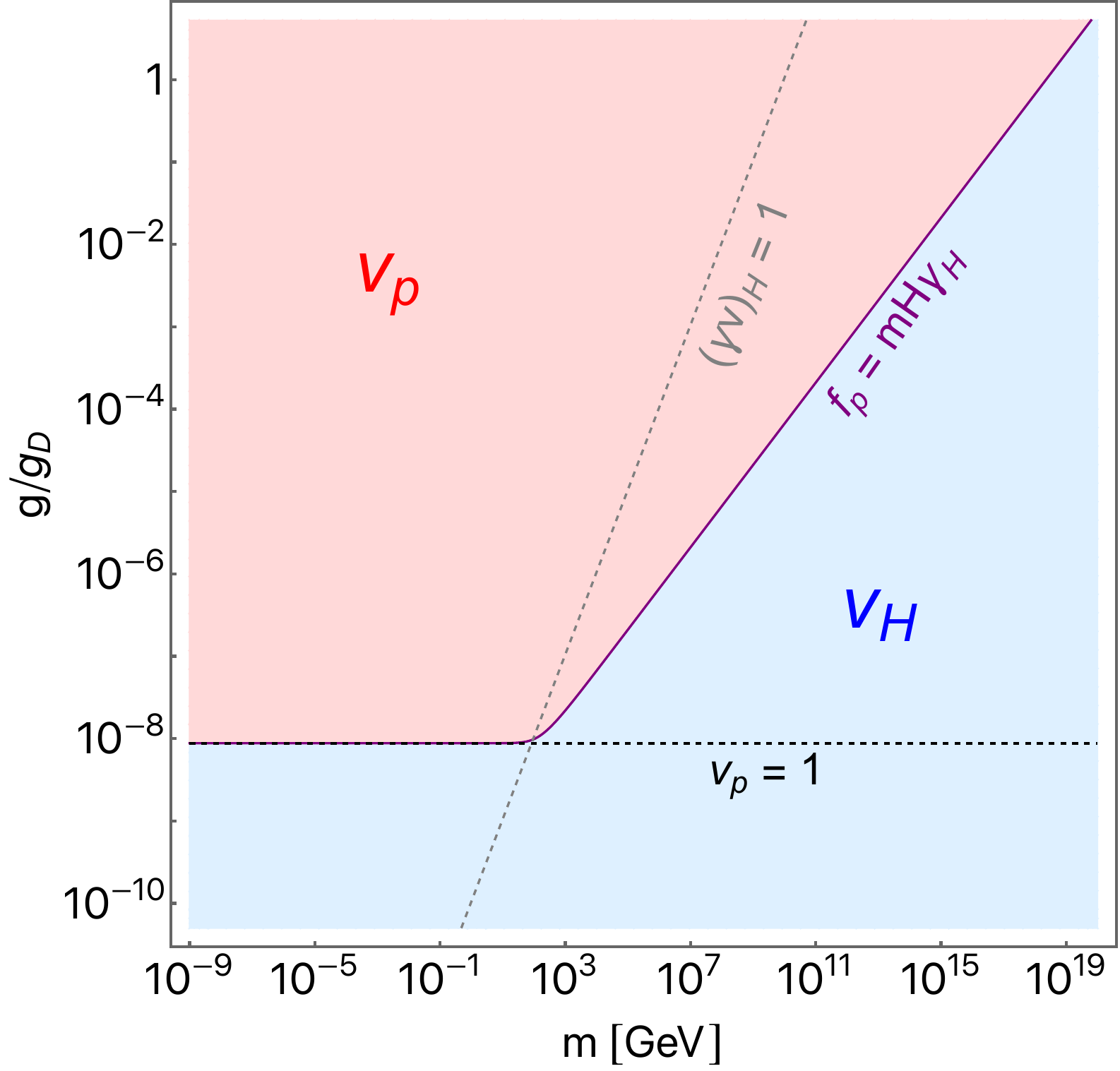}
\caption{Velocity of monopoles accelerated by primordial magnetic fields during radiation domination.
The solid purple curve shows the combination of the monopole mass and charge for which the friction term due to the interaction with the primordial plasma, $f_{\mathrm{p}}$, is comparable to the friction term due to the expansion of the universe, $m H \gamma_{\mathrm{H}}$. In the red region above the curve the monopole velocity is controlled by the plasma friction, $f_{\mathrm{p}} \gg m H \gamma_{\mathrm{H}}$. In the blue region below the curve the velocity is controlled by the Hubble friction, $f_{\mathrm{p}} \ll m H \gamma_{\mathrm{H}}$. The dashed gray line shows where $(\gamma v)_{\mathrm{H}} = 1$. The dashed black horizontal line shows where $v_{\mathrm{p}} = 1$. Here we assume $B_0 = 10^{-15}~\mathrm{G}$ and $g_{*} = \mathcal{N}_c = 10.75$.}
\label{fig:ChargeMass}
\end{figure}

Whether the monopole velocity during radiation domination follows $v_{\ro{p}}$ or $v_{\ro{H}}$ depends on the monopole properties and the magnetic field strength. 
This is illustrated in Figure~\ref{fig:ChargeMass} in the $m$-$g$ plane, where we took the field strength such that it becomes
$B_0 = 10^{-15}~\mathrm{G}$ today.
The numbers of relativistic (charged) degrees of freedom are fixed to $g_{*} = \mathcal{N}_c = 10.75$.
The purple curve shows where the plasma and Hubble frictions in the monopole's equation of motion Eq.~\eqref{eq:eqVelocity} are comparable, i.e. $f_{\mathrm{p}} = m H \gamma_{\mathrm{H}}$. 
In the red region the plasma friction is dominant ($f_{\mathrm{p}} \gg m H \gamma_{\ro{H}}$) and the monopole velocity is given by $v_{\ro{p}}$.
On the other hand, in the blue region the Hubble friction is dominant ($f_{\mathrm{p}} \ll m H \gamma_{\ro{H}}$) and the velocity is given by $v_{\ro{H}}$.

The balance condition $f_{\ro{p}} = m H \gamma_{\ro{H}}$ is rewritten using Eqs.~(\ref{effep}) and (\ref{eq:vHubble}) as
\begin{equation}
\label{eq:balance}
 \left(\frac{e^2 g^2 \mathcal{N}_c T^2}{16 \pi^2} \right)^2
\sim (g B)^2 + (m H)^2 .
\end{equation}
This can be solved for the magnetic charge, and the solution $g = g_{\ro{min}}$ is approximated by:
\begin{equation}
\label{eq:3.10}
g_{\ro{min}} \sim 
 \begin{dcases}
\frac{16 \pi^2}{e^2 \mathcal{N}_c } \frac{B}{T^2} 
  & \mathrm{for}\, \, \,  
m \ll \frac{16 \pi^2}{e^2 \mathcal{N}_c}\frac{B^2}{H T^2},
\\
\left( \frac{16 \pi^2}{e^2 \mathcal{N}_c } \frac{m H}{T^2}  \right)^{1/2}
  & \mathrm{for}\, \, \,  
m \gg \frac{16 \pi^2}{e^2 \mathcal{N}_c}\frac{B^2}{H T^2}.
 \end{dcases}
\end{equation}
For $g > g_{\ro{min}}$ the monopole velocity approaches~$v_{\ro{p}}$,
and for $g < g_{\ro{min}}$ it approaches~$v_{\ro{H}}$.
In other words, $g_{\ro{min}}$ sets the minimum charge for a monopole during radiation domination to lose its kinetic energy mainly through its interaction with the plasma.
The expressions of Eq.~(\ref{eq:3.10}) describe the two asymptotic behaviors of the purple curve in the figure. 
In the first line the balance condition is realized for relativistic velocities ($(\gamma v)_{\ro{H}} \gg 1$), while the second line is for nonrelativistic velocities ($(\gamma v)_{\ro{H}} \ll 1$).
In the figure, the dashed gray line shows where $(\gamma v)_{\mathrm{H}} = 1$, with $(\gamma v)_{\mathrm{H}} < 1$ on its right side.
One actually sees that the dashed gray and purple lines intersect at the point where the purple line bends.
We also note that the first line of Eq.~(\ref{eq:3.10}) corresponds to the charge that gives $v_{\ro{p}} = 1$; this is depicted in the figure by the dashed black line. In the region below this line the expression Eq.~\eqref{eq:vPlasma} yields $v_{\ro{p}} > 1$, indicating that the plasma friction does not yield a terminal velocity for monopoles.

The expressions in Eq.~(\ref{eq:3.10}) are time independent during radiation, up to mild variations due to the change in the numbers of relativistic degrees of freedom. 
The cosmic temperature and the Hubble rate during the radiation-dominated epoch are related to the redshift as
\begin{equation}
 T \sim 1\, \ro{MeV} \left(\frac{10^{-10}}{a/a_0}\right),
\quad
 H \sim 10^{-15}\, \ro{eV} \left(\frac{10^{-10}}{a/a_0}\right)^2,
\label{eq:TH_RD}
\end{equation}
where we ignored their mild dependence on $g_{*(s)}$.
For the number of relativistic charged degrees of freedom, hereafter 
we use $\mathcal{N}_{c} \sim 10$ as a reference value. Combining these  with the magnetic scaling $B = B_0 (a_0/a)^2$, one can rewrite Eq.~(\ref{eq:3.10}) as
\begin{equation}
\label{eq:gmin}
g_{\ro{min}} \sim 
  \begin{dcases}
10^{-8} g_{\ro{D}} \left( \frac{B_0}{10^{-15}~\mathrm{G}} \right)
  & \mathrm{for}\, \, \,  
m \ll 10^{2}\, \mathrm{GeV} \left( \frac{B_0}{10^{-15}\, \mathrm{G}} \right)^2,
\\
10^{-1} g_{\ro{D}} \left(\frac{m}{10^{17} \, \ro{GeV}} \right)^{1/2}
  & \mathrm{for}\, \, \,  
m \gg 10^{2}\, \mathrm{GeV} \left( \frac{B_0}{10^{-15}\, \mathrm{G}} \right)^2.
 \end{dcases}
\end{equation}
The terminal velocities of Eqs.~(\ref{effep}) and (\ref{eq:vHubble}) can also be rewritten as,
\begin{gather}
\label{eq:vPlasmaComp}
    v_{\mathrm{p}} \sim 10^{-8} \left( \frac{g}{g_{\mathrm{D}}} \right)^{-1} \left( \frac{B_0}{10^{-15}~\mathrm{G}} \right),
\\
\label{eq:vHubbleComp}
    \left( \gamma v \right)_{\mathrm{H}} \sim 10^{-7} \left( \frac{m}{10^{17}~\mathrm{GeV}} \right)^{-1} \left( \frac{g}{g_{\mathrm{D}}} \right) \left( \frac{B_0}{10^{-15}~\mathrm{G}} \right) .
\end{gather}

The monopole velocity during radiation domination shown in the right parts of Figure~\ref{fig:VelRatio} can be understood from Figure~\ref{fig:ChargeMass}, and by noting that the terminal velocities scale with the monopole mass and charge as 
$v_{\ro{p}} \propto g^{-1}$, $(\gamma v)_{\ro{H}} \propto g m^{-1}$.
The variation of the velocities in Figure~\ref{fig:velocity} is understood by moving horizontally in Figure~\ref{fig:ChargeMass} along $g = 10^{-3} g_{\ro{D}}$; for small~$m$ the velocity is set to $v_{\ro{p}}$ which is independent of~$m$ (cf. purple and blue lines in Figure~\ref{fig:velocity}), while for large~$m$ the velocity is $v_{\ro{H}}$ which decreases with~$m$ (cf. green, orange, and red lines).
On the other hand, Figure~\ref{fig:velocity2} corresponds to moving vertically in Figure~\ref{fig:ChargeMass} along $m = 10^{11}\, \ro{GeV}$; for small~$g$ the velocity~$v_{\ro{H}}$ increases with~$g$ (cf. purple, blue, and green lines in Figure~\ref{fig:velocity2}), while for large~$g$ the velocity~$v_{\ro{p}}$ decreases with~$g$ (cf. orange and red lines).

With constant monopole velocities, the dissipation rate ratio grows as $\Pi_{\mathrm{acc}} / \Pi_{\mathrm{red}} \propto a$.
Thus, requiring negligible monopole backreaction while there are abundant charged particles in the universe amounts to 
demanding that this ratio is smaller than unity at $e^+ e^-$ annihilation, i.e.,
\begin{equation}
 \left(
\frac{\Pi_{\mathrm{acc}} }{\Pi_{\mathrm{red}}}
 \right)_{T \sim 1 \, \mathrm{MeV}} < 1.
\label{eq:PiPi1MeV}
\end{equation}
This also means that the bounds we derive in this subsection apply as long as the primordial magnetic fields have been generated before $e^+ e^-$ annihilation.

In the case of $g > g_{\ro{min}}$, the ratio $\Pi_{\mathrm{acc}} / \Pi_{\mathrm{red}}$ is evaluated by substituting $v = v_{\ro{p}}$ into Eq. (\ref{eq:ratio}). 
Hence the condition in Eq.~(\ref{eq:PiPi1MeV})
can be rewritten by using $n = n_0 (a_0/a)^3$ as an upper bound on the present-day monopole number density,
\begin{equation}
    n_0 \lesssim 
    10^{-21}~\mathrm{cm}^{-3}.
\end{equation}
We express this condition also in terms of the present-day monopole flux $F = n_0 v_0 / 4 \pi$:
\begin{equation}
\label{eq:BoundDuring}
    F \lesssim 
    10^{-14}~\mathrm{cm}^{-2} \mathrm{sr}^{-1} \mathrm{s}^{-1} \left( \frac{v_0}{10^{-3}} \right) . 
\end{equation}
The bound is mainly determined by the temperature and redshift at $e^+ e^-$ annihilation, and thus is independent of the amplitude of the magnetic fields and of the mass and the charge of the monopoles.
However, the red region in Figure~\ref{fig:ChargeMass} where the bound can be applied ($g > g_{\ro{min}}$) becomes smaller for stronger magnetic fields.

In the case of $g < g_{\ro{min}}$, the monopoles do not efficiently transfer the magnetic energy to the plasma. 
Hence their presence does not lead to the dissipation of primordial magnetic fields, but can only induce oscillations of the fields and 
affect their redshift evolution.
In order for the fields' redshifting to be unaltered by monopoles, 
the condition $\Pi_{\mathrm{acc}} / \Pi_{\mathrm{red}} < 1$ should hold all the way until today.
However, in order to connect with the bounds we derived for
$g > g_{\ro{min}}$, here let us only require the redshifting to be unaltered at temperatures $T > 1\, \ro{MeV}$ and 
impose Eq.~(\ref{eq:PiPi1MeV}).
We further limit our analysis to nonrelativistic monopoles, 
i.e. $(\gamma v)_{\ro{H}} \lesssim 1$,
which from Eq.~(\ref{eq:vHubbleComp}) is equivalent to considering masses of:
\begin{equation}
 m \gtrsim 10^{10}\, \ro{GeV} \left(\frac{g}{g_{\ro{D}}}\right)
\left(\frac{B_0}{10^{-15}\, \ro{G}}\right).
\label{eq:nonrel_mass}
\end{equation}
(We are thus focusing on the region in Figure~\ref{fig:ChargeMass} on the right of both the purple and gray dashed lines.)
Then the dissipation rate ratio can be evaluated
by substituting $v = v_{\ro{H}} \sim gB / m H$ into Eq.~(\ref{eq:ratio}), 
and the condition of Eq.~(\ref{eq:PiPi1MeV}) translates into:
\begin{equation}
    n_0 \lesssim 
    10^{-22} \, \mathrm{cm}^{-3}  \left( \frac{m}{10^{17}\, \mathrm{GeV}} \right) \left( \frac{g}{g_{\mathrm{D}}} \right)^{-2} .
\label{eq:3.19}
\end{equation}
The condition in terms of the present-day monopole flux is:
\begin{equation}
\label{eq:BoundDuring2}
    F \lesssim 10^{-16} \, \mathrm{cm}^{-2} \mathrm{sr}^{-1} \mathrm{s}^{-1} 
\left( \frac{m}{10^{17} \, \mathrm{GeV}} \right) \left( \frac{g}{g_{\mathrm{D}}} \right)^{-2} 
\left( \frac{v_0}{10^{-3}} \right). 
\end{equation}

\subsection{Reheating epoch}
\label{sec:vel:before}

During reheating the universe is effectively matter-dominated, and the Hubble rate redshifts as $H \propto a^{-3/2}$.
The final results of this section depend only mildly on the numbers of relativistic degrees of freedom. Thus, for simplicity we ignore their time dependences and use $g_{*(s)} \sim \mathcal{N}_c \sim 100$ in the following analyses. 
We also assume the plasma particles to be in thermal equilibrium during reheating. Under these assumptions, the temperature of the primordial plasma redshifts as $T \propto a^{-3/8}$ \cite{Kolb:1990vq}.
Consequently, the plasma-induced terminal velocity scales as $v_{\mathrm{p}} \propto a^{-5/4}$, and the Hubble-induced terminal velocity scales as $(\gamma v)_{\mathrm{H}} \propto a^{-1/2}$; 
the redshifting of the former being faster is related to the fact that the monopole velocity during reheating can make a transition from $v_{\ro{H}}$ to $v_{\ro{p}}$ but not vice versa~\cite{Kobayashi:2022qpl}, as was shown in Figure~\ref{fig:VelRatio}.
Combining this with the discussion in the previous subsection, we see that monopoles make the transition to the $v_{\ro{p}}$-branch before reheating completes if the charge satisfies 
$g > g_{\ro{min}}$, with $g_{\ro{min}}$ given in Eq.~\eqref{eq:gmin}.

For the case of $g > g_{\mathrm{min}}$, we define $t_*$ as the time when the plasma friction takes over the Hubble friction,
i.e. $f_{\mathrm{p} *} = m H_* \gamma_{\ro{H} *}$ (the subscript ``$*$'' stands for quantities computed at time $t_*$).
For $t < t_*$ the monopoles move at the terminal velocity $v_{\mathrm{H}}$, while for $t > t_*$ the monopoles follow $v_{\mathrm{p}}$. 
The balance of the frictional forces is written as Eq.~\eqref{eq:balance}, which can be transformed into an equation for $H_*$ by considering that $f_{\mathrm{p}} \propto H^{1/2}$ and $B \propto H^{4/3}$ during reheating as:
\begin{equation}
\label{eqHstar}
f_{\ro{p \, dom}}^2 \sim 
 (g B_{\mathrm{dom}})^2 \left( \frac{H_*}{H_{\mathrm{dom}}} \right)^{5/3} + (m H_{\mathrm{dom}})^2 \left( \frac{H_*}{H_{\mathrm{dom}}} \right).
\end{equation}
Considering that the right-hand side is dominated by one of the terms, this equation can be solved approximately as
\begin{equation}
\label{eq:H_star}
    H_* \sim 
    \begin{dcases}  
    H_{\mathrm{dom}} \left( \frac{f_{\mathrm{p}}}{g B}  \right)_{\mathrm{dom}}^{6/5} 
    & \mathrm{for}\, \, \,  m \ll \bar{m},
    \\
    H_{\mathrm{dom}} \left( \frac{f_{\mathrm{p}}}{m H} \right)_{\mathrm{dom}}^2
    & \mathrm{for}\, \, \,  m \gg \bar{m}.
    \end{dcases}
\end{equation}
Here $\bar{m}$ is defined as
\begin{equation}
 \bar{m} = \left(
\frac{g^{3} B^{3} f_{\mathrm{p}}^{2}}{H^5}
\right)_{\mathrm{dom}}^{1/5},
\end{equation}
however we note that the combination $ B^{3} f_{\mathrm{p}}^{2} / H^5$ is actually time-independent during the reheating epoch, up to mild variations from changes in the numbers of relativistic degrees of freedom.
Using the relations in Eq.~\eqref{eq:TH_RD} computed at $t_{\mathrm{dom}}$, we can rewrite these expressions as
\begin{gather}
\label{Hstar}
    H_* \sim 
    \begin{dcases}  
10^{5}\, \mathrm{GeV}  \left( \frac{g}{g_{\mathrm{D}}}\right)^{6/5} 
\left( \frac{B_0}{10^{-15} \, \mathrm{G}} \right)^{-6/5}
\left( \frac{T_{\mathrm{dom}}}{10^{6}\, \mathrm{GeV}} \right)^2 
    & \mathrm{for}\, \, \,  m \ll \bar{m},
    \\
10^{-1} \, \mathrm{GeV}  
\left( \frac{m}{10^{17}~\mathrm{GeV}} \right)^{-2}
\left( \frac{g}{g_{\mathrm{D}}} \right)^4 \left( \frac{T_{\mathrm{dom}}}{10^{6}\, \mathrm{GeV}} \right)^2 
    & \mathrm{for}\, \, \,  m \gg \bar{m},
    \end{dcases}
\\
\label{eq:barM}
    \bar{m} \sim 10^{14} \, \mathrm{GeV}  
\left( \frac{g}{g_{\mathrm{D}}} \right)^{7/5}
\left( \frac{B_0}{10^{-15}\, \mathrm{G}} \right)^{3/5} .
\end{gather}
If $m \ll \bar{m}$, the balancing of the frictional forces happens 
while the monopoles are relativistic, and the monopole velocity jumps from an ultrarelativistic~$v_{\ro{H}}$ to a mildly relativistic~$v_{\ro{p}}$. 
On the other hand if  $m \gg \bar{m}$, the balancing happens while the monopoles are nonrelativistic, and the velocity transition is smooth.
These explain the behavior of the monopole velocity described at the end of Section~\ref{sec:3.1}.

The relation $H_* > H_{\ro{dom}}$ holds if $g > g_{\ro{min}}$. While monopoles carrying such a charge move at the velocity~$v_{\mathrm{p}}$ and transfer the magnetic field energy into the plasma, the dissipation ratio decreases in time as 
$\Pi_{\mathrm{acc}} / \Pi_{\mathrm{red}} \propto a^{-3/4}$
(this behavior is shown in Figure~\ref{fig:ratio} by the middle parts of the blue and purple lines).
Hence requiring that the fields survive during reheating amounts to imposing 
\begin{equation}
 \left(
\frac{\Pi_{\mathrm{acc}} }{\Pi_{\mathrm{red}}}
 \right)_* < 1.
\label{eq:PiPistar}
\end{equation}
Combining this with $v = v_{\ro{p}}$ and Eq.~(\ref{Hstar})
yields a bound on the monopole abundance,\footnote{For $m \ll \bar{m}$, the ratio $\Pi_{\mathrm{acc}} / \Pi_{\mathrm{red}}$ undergoes a jump shortly after~$t_*$ (cf. purple line in Figure~\ref{fig:ratio}). By plugging $v = v_{\ro{p}}$, here we are approximately evaluating the value of $\Pi_{\mathrm{acc}} / \Pi_{\mathrm{red}}$ after the jump.} 
\begin{equation}
\label{eq:nperri}
\begin{split}
    n_0 \lesssim
    \ro{max.} \Biggl\{
    &10^{-16} \, \mathrm{cm}^{-3} 
\left( \frac{g}{g_{\mathrm{D}}} \right)^{-3/5}
\left( \frac{B_0}{10^{-15}\, \mathrm{G}} \right)^{3/5} \left( \frac{T_{\mathrm{dom}}}{10^{6}\, \mathrm{GeV}} \right) ,
    \\
    &10^{-13}\, \mathrm{cm}^{-3} \left( \frac{m}{10^{17}\, \mathrm{GeV}} \right) 
\left( \frac{g}{g_{\mathrm{D}}} \right)^{-2}
\left( \frac{T_{\mathrm{dom}}}{10^{6}\, \mathrm{GeV}} \right) \Biggr\} .
\end{split}
\end{equation}
The first (second) line sets the condition when $m$ is smaller (larger) than $\bar{m}$ given in Eq.~(\ref{eq:barM}).
In terms of the present monopole flux, the condition is:
\begin{equation}
\label{eq:perri}
\begin{split}
    F \lesssim
    \ro{max.} \Biggl\{
    &10^{-10} \, \mathrm{cm}^{-2}\mathrm{sr}^{-1} \mathrm{s}^{-1}  
\left( \frac{g}{g_{\mathrm{D}}} \right)^{-3/5}
\left( \frac{B_0}{10^{-15}\ \mathrm{G}} \right)^{3/5} \left( \frac{T_{\mathrm{dom}}}{10^{6}\, \mathrm{GeV}} \right) 
\left( \frac{v_0}{10^{-3}} \right) , 
    \\
    &10^{-7} \, \mathrm{cm}^{-2} \mathrm{sr}^{-1} \mathrm{s}^{-1}  
\left( \frac{g}{g_{\mathrm{D}}} \right)^{-2} 
\left( \frac{m}{10^{17} \, \mathrm{GeV}} \right) \left( \frac{T_{\mathrm{dom}}}{10^{6}\, \mathrm{GeV}} \right) 
\left( \frac{v_0}{10^{-3}} \right) 
\Biggr\} .
\end{split}
\end{equation}

The bounds in Eqs. (\ref{eq:nperri}) and (\ref{eq:perri}) assume that the velocity transition happens during the reheating epoch, and in particular after the primordial magnetic fields are generated, hence
\begin{equation}
 H_* < H_{\ro{end}}, \, H_{\ro{inf}}.
\label{eq:HH}
\end{equation}
Here $H_{\ro{inf}}$ is the inflationary Hubble rate, which is constrained 
by current observational limits on primordial gravitational waves as~\cite{BICEP:2021xfz}
\begin{equation}
H_{\ro{inf}} \lesssim 10^{14}\, \ro{GeV}. 
\label{eq:inf_bound}
\end{equation}
Thus the condition of Eq.~(\ref{eq:HH}) would be violated if $H_*$ becomes very large, for instance due to a large $T_{\ro{dom}}$.
We also note that, going back in time in the reheating epoch, the magnetic energy grows relative to the total density as $\rho_B / \rho_{\mathrm{tot}} \propto a^{-1}$. 
For magnetic fields generated at the end of inflation or during reheating, requiring that they have never dominated the universe yields a constraint on the scale of magnetogenesis as
\begin{equation}
 H_{\mathrm{end}} \lesssim 10^{22}\, \mathrm{GeV}
\left( \frac{T_{\mathrm{dom}}}{10^6\, \mathrm{GeV}} \right)^2
\left(\frac{10^{-15}\, \mathrm{G}}{B_0}\right)^3.
\label{eq:2.28}
\end{equation}
For $B_0 \sim 10^{-15}\, \ro{G}$ and $g \lesssim g_{\ro{D}}$, there is a wide range for $H_{\ro{end}}$ where both this and (\ref{eq:HH}) are satisfied.
However this constraint can become relevant for magnetic black holes, as we will later see.
If the fields are generated after $t_*$, then the monopole bound becomes weaker; such cases are studied in~\cite{Kobayashi:2022qpl}.

If the charge is as small as $g < g_{\mathrm{min}}$, the monopole velocity never approaches~$v_{\mathrm{p}}$.
We can also derive the condition for such monopoles not to affect the redshifting of the magnetic fields during the reheating epoch. 
In this case the dissipation rate ratio is non-decreasing during reheating:
It increases as $\Pi_{\mathrm{acc}} / \Pi_{\mathrm{red}} \propto a^{1/2}$ while the monopoles move at relativistic~$v_{\ro{H}}$, 
then stays constant after $v_{\ro{H}}$ becomes nonrelativistic. 
Hence we impose the condition at the onset of radiation domination,
\begin{equation}
 \left(
\frac{\Pi_{\mathrm{acc}} }{\Pi_{\mathrm{red}}}
 \right)_{\ro{dom}} < 1.
\label{eq:PiPidom}
\end{equation}
This condition assumes that the magnetic fields were produced before
the radiation-dominated epoch begins. 
We further focus on monopoles that become nonrelativistic before~$t_{\ro{dom}}$, which amounts to considering masses satisfying the condition of Eq.~(\ref{eq:nonrel_mass}).
This allows us to plug $v_{\ro{dom}} \sim (gB / m H)_{\ro{dom}}$ into Eq.~(\ref{eq:PiPidom}). One can check that the upper bound on the present-day monopole number density thus obtained is the same as the second line of Eq.~(\ref{eq:nperri}), and the flux bound is the same as the second line of Eq.~(\ref{eq:perri}).

\subsection{Summary of bounds from primordial magnetic fields}
\label{sec:summPrim}

We have seen that the bounds on the monopole flux from the survival of primordial magnetic fields are described by Eq.~\eqref{eq:BoundDuring} during radiation domination, and by Eq.~\eqref{eq:perri} during reheating. The bounds are valid under the condition $g > g_{\mathrm{min}}$, where the minimum magnetic charge $g_{\mathrm{min}}$ is given in Eq.~\eqref{eq:gmin}.
The bound from radiation domination assumes that primordial magnetic fields are generated before $e^+ e^-$ annihilation.
For the bound from reheating it is further assumed that the scales of magnetogenesis~$H_{\ro{end}}$, inflation~$H_{\ro{inf}}$, and $H_*$ given in Eq.~(\ref{Hstar}) satisfy the condition of Eq.~(\ref{eq:HH}); 
here $H_{\ro{end}}$ is also constrained by Eq.~(\ref{eq:2.28}),
and $H_{\ro{inf}}$ by Eq.~(\ref{eq:inf_bound}).

The flux bound from radiation domination is independent of the amplitude of the magnetic fields and of the mass and the charge of the monopoles, although the minimum charge~$g_{\mathrm{min}}$ depends on the field strength and mass.
The bound from reheating depends on a number of parameters, and in particular it becomes stronger for larger charges and lower reheating temperatures.

For monopoles with $g < g_{\mathrm{min}}$, we derived conditions for them not to alter the redshifting of primordial magnetic fields.\footnote{Monopoles with $g > g_{\mathrm{min}}$ can also affect the magnetic redshifting, however we did not analyze this case.}
Focusing on masses satisfying Eq.~(\ref{eq:nonrel_mass}), 
the condition during radiation domination gives the flux bound in Eq.~(\ref{eq:BoundDuring2}) (assuming magnetogenesis before $e^+ e^-$ annihilation), and 
from reheating arises the bound which takes the same expression as
the second line of Eq.~(\ref{eq:perri}) 
(assuming magnetogenesis before the radiation-dominated epoch begins).
For these bounds, we stress that primordial magnetic fields can survive even if they are violated, however in such cases one needs to take into account the monopoles in order to assess the cosmological evolution of primordial magnetic fields.

\begin{figure}[!t]
  \centering
  \includegraphics[width=0.9\textwidth]{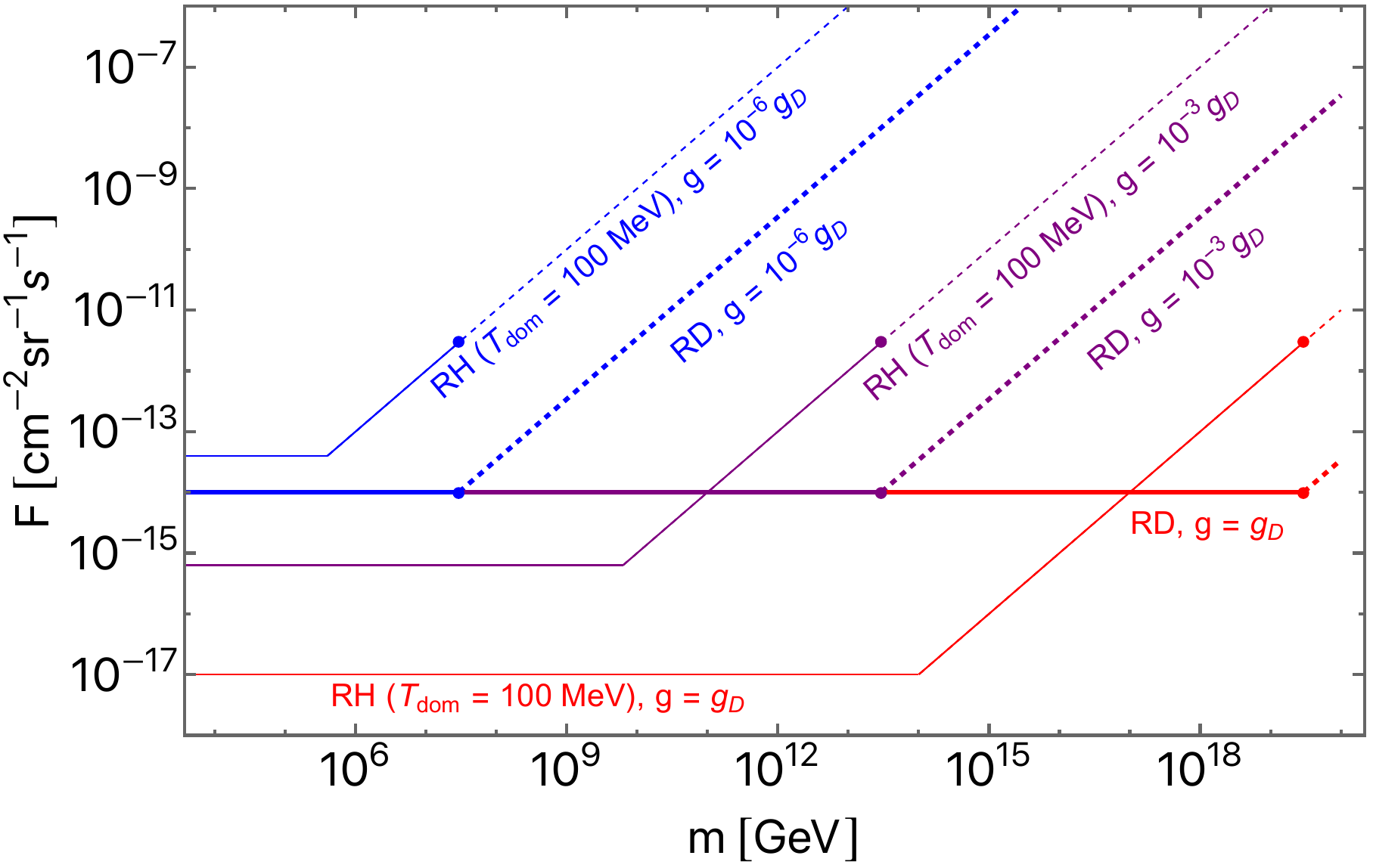}
\caption{Upper bounds on the monopole flux as a function of mass, from the survival of primordial magnetic fields during radiation domination (thick lines) and reheating era (thin lines). The magnetic charge is varied as 
$g = g_{\ro{D}}$ (red),
$10^{-3} g_{\ro{D}}$ (purple),
$10^{-6} g_{\ro{D}}$ (blue).
The dotted parts of the lines show where the fields exhibit modified redshifting behaviors, instead of being dissipated.
The magnetic field strength is taken such that it realizes a present-day value of $B_0 = 10^{-15}\, \mathrm{G}$, and the monopole velocity today is fixed to $v_0 = 10^{-3}$.
The reheating bounds assume a reheating temperature of $T_{\mathrm{dom}} =100\, \mathrm{MeV}$.
}
\label{fig:Primordial}
\end{figure}

In Figure~\ref{fig:Primordial} we show the upper bounds on the monopole flux from radiation domination (thick lines), and from reheating with $T_{\mathrm{dom}} = 100\, \mathrm{MeV}$ (thin lines). The magnetic charge is varied as $g = g_{\mathrm{D}}$ (red), $g = 10^{-3} g_{\mathrm{D}}$ (purple), and $g = 10^{-6} g_{\mathrm{D}}$ (blue). 
We have chosen a rather low reheating temperature just a few orders of magnitude above the scale of Big Bang Nucleosynthesis, as an optimal value for the reheating bound.
The purple, blue, and red thick lines overlap in the left part of the plot.
Here we assume $v_0 = 10^{-3}$, and $B_0 = 10^{-15}\, \mathrm{G}$. 
The solid parts of the lines are based on the survival of primordial fields ($g > g_{\ro{min}}$), while the dashed parts are from the requirement that the redshifting of the primordial fields is unaltered ($g < g_{\ro{min}}$). 
For the masses shown in the plot, $g_{\mathrm{min}}$ is given by the second line of \eqref{eq:gmin}, and thus the condition $g > g_{\mathrm{min}}$ can be rewritten as:
\begin{equation}
\label{eq:purple}
   m \lesssim 10^{19}~\mathrm{GeV} \left( \frac{g}{g_{\mathrm{D}}} \right)^2 .
\end{equation}
The points in the plot show where this bound is saturated.
We also note that the parameters used for the plot allow for ranges of values for $H_{\ro{end}}$ and $H_{\ro{inf}}$ that satisfy the assumptions in Eqs.~(\ref{eq:HH}), (\ref{eq:inf_bound}), and (\ref{eq:2.28}). 
Moreover, the condition in Eq.~\eqref{eq:nonrel_mass} is satisfied on the dashed lines.

As shown in the plot, for $T_{\mathrm{dom}} = 100\, \mathrm{MeV}$ the bound from reheating is stronger than the bound from radiation domination at low masses, for $g \gtrsim 10^{-5} g_{\mathrm{D}}$. 
However for $T_{\mathrm{dom}} \gtrsim 10^2\, \ro{GeV}$, the bound from radiation domination becomes stronger than the bound from reheating even at $g = g_{\mathrm{D}}$. 
We stress again that the bound from the survival of primordial fields during radiation domination does not weaken for smaller charges (although its range of applicability shrinks to smaller masses); this feature makes the radiation domination bound particularly useful for constraining minicharged monopoles.

\section{Comparison of bounds}
\label{sec:comparison}

\begin{figure}[!t]
  \centering
  \begin{subfigure}[b]{0.496\textwidth}
    \centering
    \includegraphics[width=\textwidth]{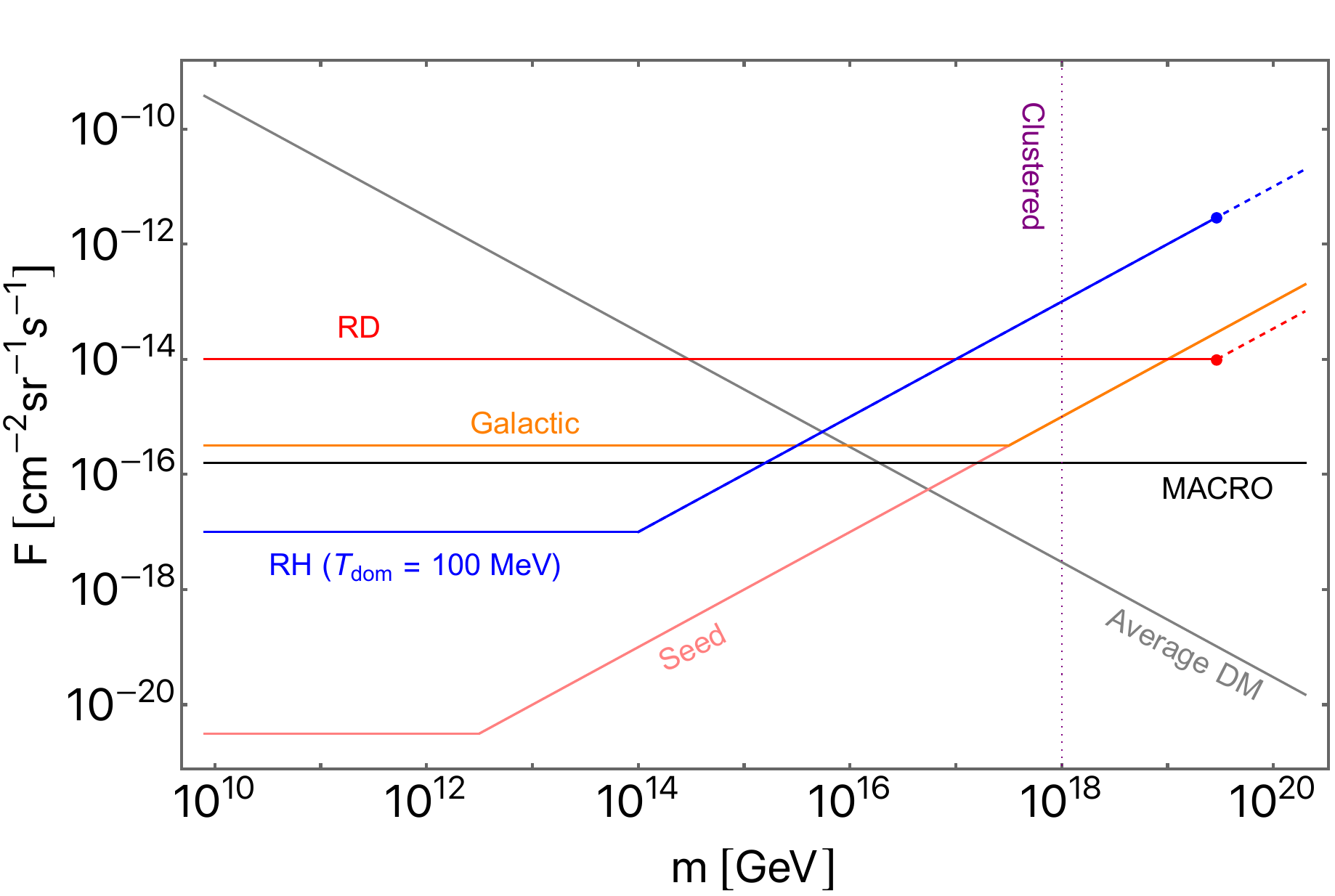}
    \caption{$g = g_{\mathrm{D}}$}
    \label{fig:dirac}
  \end{subfigure}
  \hfill
  \begin{subfigure}[b]{0.496\textwidth}
    \centering
    \includegraphics[width=\textwidth]{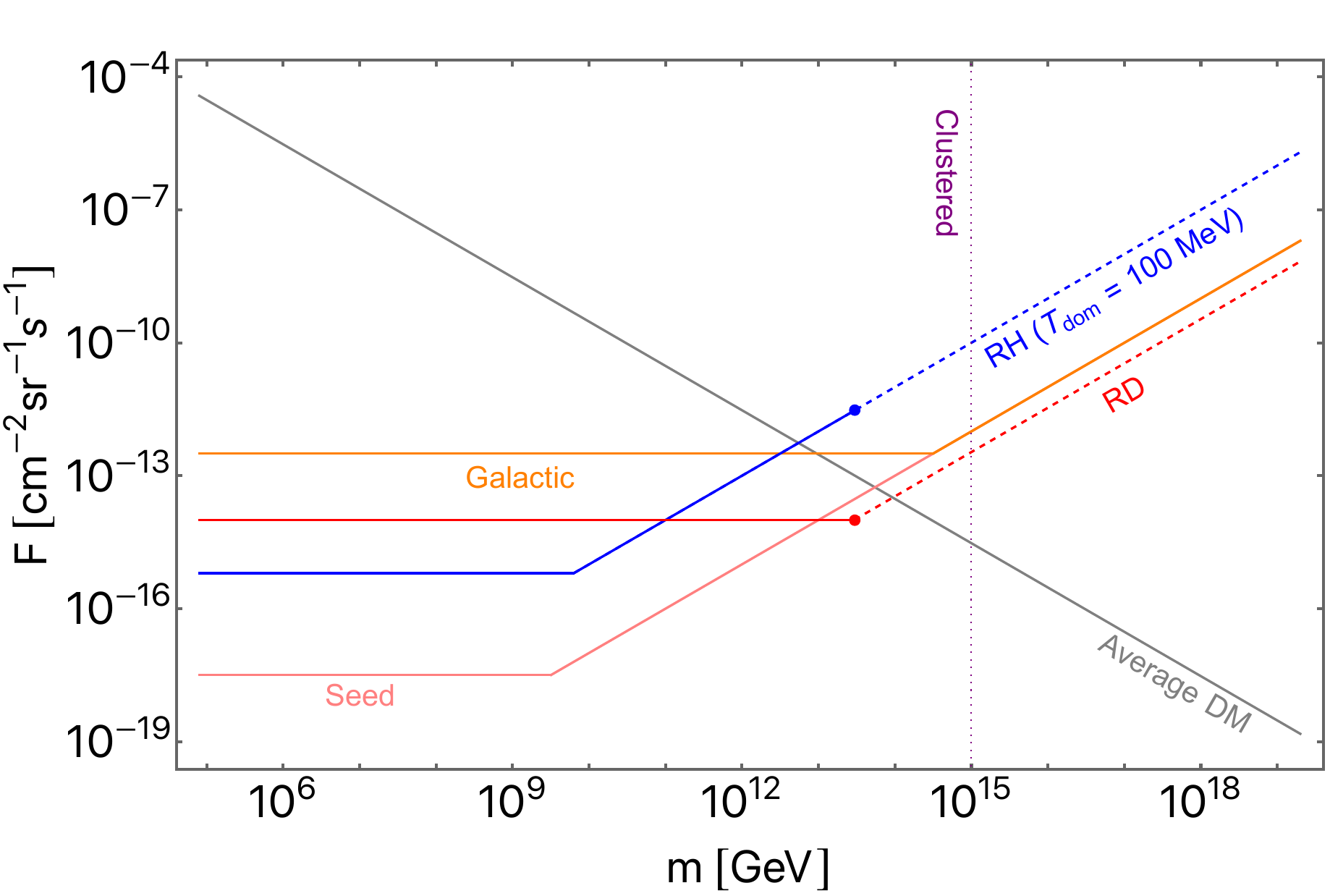}
    \caption{$g = 10^{-3} g_{\mathrm{D}}$}
    \label{fig:10m2}
  \end{subfigure}
  \hfill
  \begin{subfigure}[b]{0.496\textwidth}
    \centering
    \includegraphics[width=\textwidth]{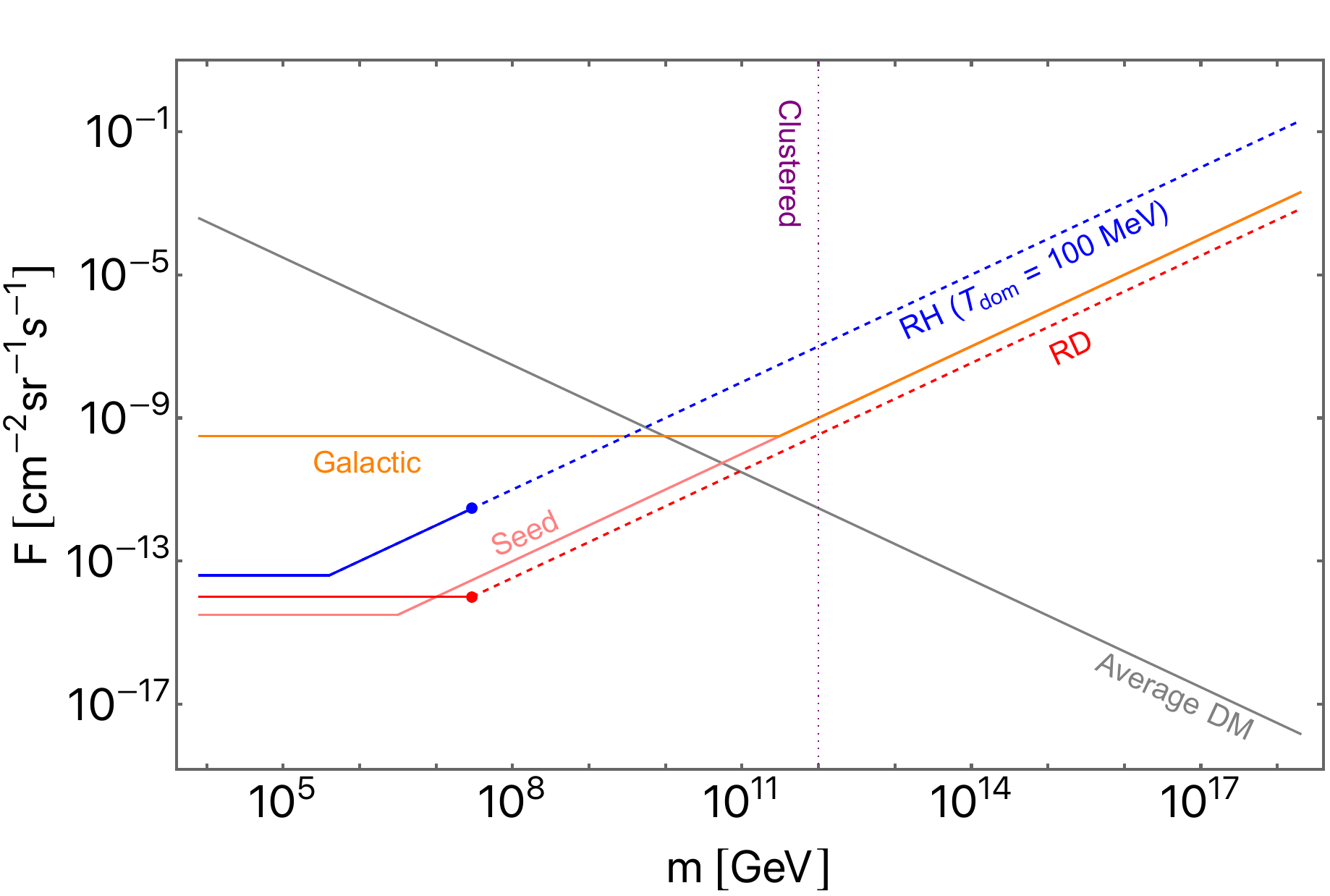}
    \caption{$g = 10^{-6} g_{\mathrm{D}}$}
    \label{fig:10m5}
  \end{subfigure}
  \hfill
  \begin{subfigure}[b]{0.496\textwidth}
    \centering
    \includegraphics[width=\textwidth]{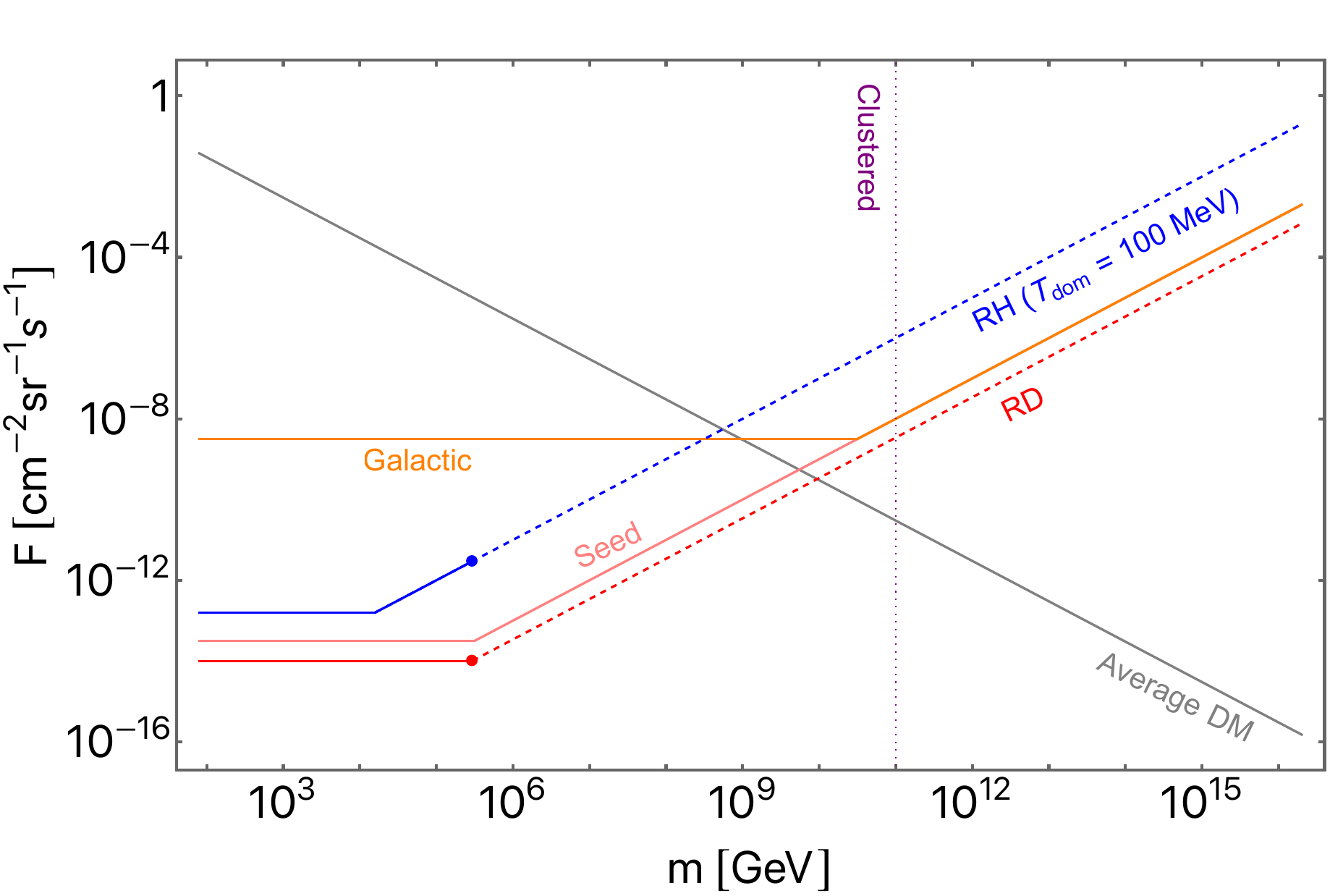}
    \caption{$g = 10^{-7} g_{\mathrm{D}}$}
    \label{fig:10m6}
  \end{subfigure}
  \caption{Upper bounds on the monopole flux as functions of mass, for
different values of magnetic charge.
Gray: cosmological bound from comparison with the average dark matter density in the universe.
Orange: bound from Galactic magnetic fields of $B = 10^{-6}~\mathrm{G}$;
the lower mass limit for monopoles to stay clustered with the Galaxy is shown by the vertical dotted line.
Pink: bound from Galactic seed magnetic fields of $B = 10^{-11}~\mathrm{G}$.
Red: bound from primordial magnetic fields during radiation domination,
with present-day strength $B_0 = 10^{-15}~\mathrm{G}$.
Blue: bound from primordial magnetic fields during reheating era,
with present-day strength $B_0 = 10^{-15}~\mathrm{G}$ and reheating temperature $T_{\mathrm{dom}} = 100~\mathrm{MeV}$. 
For the primordial bounds, the dashed parts of the lines show where the fields exhibit modified redshifting behaviors, instead of being dissipated.
The panel for $g = g_{\mathrm{D}}$ also shows the limit from direct searches by the MACRO collaboration \cite{MACRO:2002jdv} in black.
We assume $v_0 = 10^{-3}$ and for the Galactic parameters $l_{\mathrm{c}} = 1~\mathrm{kpc}$, $R = 10~\mathrm{kpc}$, $\tau_{\mathrm{gen}} = 10^8~\mathrm{yr}$, and $\gamma_i - 1 = 10^{-6}$. See the text for details.
}
  \label{fig:boundComparison}
\end{figure}

Let us now compare the various bounds presented in the previous sections.
In Figure~\ref{fig:boundComparison} we show the upper bounds on the flux of magnetic monopoles as functions of the mass, for different values of the magnetic charge.
The solid gray line shows the cosmological abundance bound in Eq.~\eqref{eq:cosmological} where $\rho_{\mathrm{DM}}$ is taken as the average dark matter density in the universe, i.e. $\rho_{\mathrm{DM}} = 1.3 \times 10^{-6}~\mathrm{GeV cm^{-3}}$, along with $v_i = 10^{-3}$.
The orange line shows the bound based on the survival of Galactic magnetic fields, using Eq.~\eqref{eq:maru-j} and Galactic field strength $B = 10^{-6}\, \mathrm{G}$, along with the other parameters as $l_{\mathrm{c}} = 1~\mathrm{kpc}$, $R = 10~\mathrm{kpc}$, $\tau_{\mathrm{gen}} = 10^8~\mathrm{yr}$, and $\gamma_i - 1 = 10^{-6}$.
Using the same set of parameters, the dotted vertical line shows the lower mass limit in Eq.~\eqref{eq:maru-h} for monopoles to be clustered with our Galaxy.
The pink line shows the bound from seed Galactic fields, using again Eq.~\eqref{eq:maru-j} but with the seed field assumed to be $B = 10^{-11}\, \mathrm{G}$; the other parameters are the same as the orange line.
The pink and orange lines overlap on the right side of the plots.
The red line shows the bounds in Eqs.~\eqref{eq:BoundDuring} and (\ref{eq:BoundDuring2})) from primordial magnetic fields during radiation domination. 
The blue line shows the bound in Eq.~\eqref{eq:perri} from primordial magnetic fields during reheating, for $T_{\mathrm{dom}} = 100\, \mathrm{MeV}$.
For the primordial bounds we assume the present-day amplitude of the primordial magnetic fields to be $B_0 = 10^{-15}~\mathrm{G}$, and monopole velocity $v_0 = 10^{-3}$; moreover the solid parts of the lines are based on the survival of the fields, while the dashed parts are from the requirement that the redshifting of the fields is unaltered.
With $B_0 = 10^{-15}~\mathrm{G}$, the smallest charge with which monopoles can dissipate primordial magnetic fields is of
$10^{-8} g_{\ro{D}}$, cf. Figure~\ref{fig:ChargeMass}. 
As a value slightly above this, $g = 10^{-7} g_{\ro{D}}$ is shown in panel~\ref{fig:10m6}.
In panel~\ref{fig:dirac} for $g = g_{\ro{D}}$, we also show in black the limit from direct searches by the MACRO collaboration \cite{MACRO:2002jdv}. 

We see that monopoles with large masses are most strongly constrained by the cosmological abundance bound, while those with intermediate to low masses are mainly constrained by the Parker bounds. 
Which of the Parker bounds is most stringent for light monopoles depends on the charge.
In particular, the bound from seed Galactic magnetic fields is by far the strongest for 
monopoles with a Dirac charge, while the primordial bounds become comparable or even stronger for monopoles with small magnetic charges.
However we also note that the seed field bound further improves at very small masses if the field strength is smaller than $B = 10^{-11}\, \ro{G}$ used in the plots.

In Figure~\ref{fig:boundComparison} we have displayed the various bounds for comparison purpose. However, we should note that the target of each bound is not necessarily the same. The cosmological abundance bound and the primordial bounds constrain the average monopole density in the universe, 
while the Galactic bounds and the MACRO bound constrain the local monopole density inside the Galaxy. If monopoles are clustered with the Galaxy (although this can happen only in regions on the right of the dotted lines), their local density can be much higher than the average density; then the bounds on the local density translate into much stronger bounds on the average density.

For monopoles that can cluster with the Galaxy, namely, for masses larger than that indicated by the dotted lines, the cosmological abundance bound gives the strongest constraint. The situation is similar even when comparing with the local dark matter density;
see Figure~\ref{fig:Parker} and the discussion at the end of Section~\ref{sec:summGal}.
In other words, monopoles that cluster with our Galaxy and whose density does not exceed that of dark matter almost automatically satisfy the Parker bounds.
Hence monopoles in this mass region, if they can be produced, is a valid candidate of dark matter.

\section{Magnetically charged extremal black holes}
\label{sec:BH}

In this section we apply the bounds from the survival of magnetic fields to magnetically charged black holes. In particular, we focus on (nearly) extremal black holes for which Hawking radiation can be neglected.

At extremality the charge of a black hole is related to its mass through the relation:
\begin{equation}
\label{eq:BHrel}
 m = \sqrt{2} \, g \Mp .
\end{equation}
Extremal magnetic black holes can be considered as monopoles with large mass and small charge-to-mass ratio. Thus, all the bounds we discussed in the previous sections can basically be applied to magnetically charged black holes. 
However, the direct relation between the charge $g$ and the mass $m$ of Eq.~\eqref{eq:BHrel} changes the mass dependence of the bounds, as we will show in the following discussion.

\subsection{Bounds from galactic magnetic fields}
\label{sec:5.1}

The bound from galactic magnetic fields in Eq.~\eqref{eq:maru-j} is rewritten for extremal magnetic black holes as:
\begin{equation}
\label{eq:parkerBH}
\begin{split}
F \lesssim ~ \ro{max.} \Biggl\{
 &10^{-27}\, \ro{cm}^{-2} \ro{sec}^{-1} \ro{sr}^{-1}
\left(\frac{m}{10^{10}\, \ro{gm}} \right)^{-1}
\left( \frac{l_{\ro{c}}}{1 \, \ro{kpc}} \right)^{-1}
\left(\frac{\tau_{\ro{gen}}}{10^{8}\, \ro{yr}} \right)^{-1}
\left( \frac{ \gamma_i - 1 }{10^{-6}} \right),
\\
 &10^{-30}\, \ro{cm}^{-2} \ro{sec}^{-1} \ro{sr}^{-1}
\left(\frac{m}{10^{10}\, \ro{gm}} \right)^{-1}
\left( \frac{B}{10^{-6}\, \ro{G}} \right)
\left( \frac{R}{l_{\ro{c}}} \right)^{1/2}
\left(\frac{\tau_{\ro{gen}}}{10^{8}\, \ro{yr}} \right)^{-1}
\Biggr\},
\end{split}
\end{equation}
This bound is inversely proportional to the black hole mass.
The first (second) line sets the bound when $B$ is weaker (stronger) than the threshold value:
\begin{equation}
 \bar{B} \sim 10^{-3}\, \ro{G}
\left( \frac{l_{\ro{c}}}{1\, \ro{kpc}} \right)^{-1}
\left( \frac{l_{\ro{c}}}{R} \right)^{1/2}
\left( \frac{\gamma_i - 1}{10^{-6}} \right).
\label{eq:5.3}
\end{equation}
Since galactic fields are typically weaker than this, the bound is given by the first line, which is independent of the field strength. This implies that the bound for extremal magnetic black holes does not improve by considering seed fields.
We also remark that the conditions in Eqs.~\eqref{eq:maru-n} and~\eqref{eq:maru-z}, which are necessary for the bound to apply, can be violated for massive magnetic black holes.\footnote{If $B$ is below the threshold value~(\ref{eq:5.3}), then \eqref{eq:maru-n} gives a stronger condition than~\eqref{eq:maru-z}.}

For extremal magnetic black holes that are initially bound in a galaxy, their escape time is obtained using Eq.~\eqref{eq:tau_esc} as,
\begin{equation}
\begin{split}
    \tau_{\ro{esc}} \sim \ro{max.} 
\Biggl\{
&10^9\, \ro{yr} 
\left( \frac{B}{10^{-6}\, \ro{G}} \right)^{-1}
\left( \frac{v_{\ro{vir}}}{10^{-3}} \right),
\\
&10^{13}\, \ro{yr} 
\left( \frac{B}{10^{-6}\, \ro{G}} \right)^{-2}
\left( \frac{l_{\ro{c}}}{1 \, \ro{kpc}} \right)^{-1}
\left( \frac{v_{\ro{vir}}}{10^{-3}} \right)^3
\Biggr\} .
\end{split}
\label{eq:5.4}
\end{equation}
Note that the escape time of extremal magnetic black holes is independent of the mass, and is determined only by the galactic field properties and the virial velocity.
For galaxies similar to the Milky Way, the second line sets the escape time, which depends rather sensitively on the galactic parameters.

The work~\cite{Bai:2020spd} derived a constraint on the fraction of extremal magnetic black holes as dark matter by studying the Andromeda Galaxy, whose parameters were inferred from~\cite{Fletcher:2003ec} and 
taken as
$l_{\ro{c}} \sim 10 \, \ro{kpc}$,
$\tau_{\ro{gen}} \sim 10^{10}\, \ro{yr}$, and
$v_{\ro{vir}} \sim 10^{-3}$.
It was claimed that the large values of $l_c$ and $\tau_{\ro{gen}}$ improve the bound~(\ref{eq:parkerBH}) compared to the Milky Way.
However, these combined with the Andromeda's field strength $B \approx 5 \times 10^{-6} \, \ro{G}$~\cite{Fletcher:2003ec} yield
$\tau_{\ro{esc}} \sim 10^{10}\, \ro{yr}$, 
which is comparable to the age of Andromeda itself.
With the uncertainties in the parameters, we cannot yet give a definite answer on whether magnetic black holes can remain clustered with Andromeda until today.
However, the general lesson here is that if some galaxy appears to give a significantly stronger Parker bound on extremal magnetic black holes than the Milky Way, then it is improbable that this galaxy can currently host magnetic black holes.
The Parker bound from such a galaxy thus applies to unclustered black holes.

\subsection{Bounds from primordial magnetic fields}
 
Using the relation in Eq.~\eqref{eq:BHrel}, the terminal velocity set by the Hubble friction in Eq.~\eqref{eq:vHubble} can be rewritten for extremal magnetic black holes as:
\begin{equation}
    (\gamma v)_{\mathrm{H}} \sim \frac{B}{\sqrt{2} M_{\mathrm{pl}} H} .
\end{equation}
Thus, under the condition that the magnetic fields do not dominate the universe, i.e. $\rho_{\mathrm{B}} / \rho_{\mathrm{tot}} \sim (B/(M_{\mathrm{pl}} H))^2 \ll 1$, the velocity $v_{\mathrm{H}}$ is always nonrelativistic, $(\gamma v)_{\mathrm{H}} \ll 1$. 
From this it also follows that extremal magnetic black holes satisfy the mass bound in Eq.~(\ref{eq:nonrel_mass}).
Notice that $(\gamma v)_{\mathrm{H}}$ does not depend on the black hole mass.

With $v_{\mathrm{H}}$ being nonrelativistic, the value of $g_{\mathrm{min}}$ is given by the second line of Eq.~\eqref{eq:gmin}.
Consequently, using the relation in Eq.~\eqref{eq:BHrel} the condition $g > g_{\mathrm{min}}$ can be rewritten as:
\begin{equation}
\label{eq:BHmassLim}
    m \gtrsim 10^{-3}\, \mathrm{gm} .
\end{equation}
Extremal magnetic black holes with such masses are subject to the flux bound of Eq.~\eqref{eq:BoundDuring} which is based on the survival of primordial fields during radiation domination.

Regarding the bound from the reheating epoch,
we saw for generic monopoles that the second line of Eq.~\eqref{eq:perri} applies at larger masses for which the monopoles are nonrelativistic upon making the transition from $v_\ro{H}$ to $v_{\ro{p}}$. 
However for extremal black holes, this instead 
applies at smaller masses, 
$m < \bar{m}_{\ro{BH}}$,
with the threshold being
\begin{equation}
\label{eq:mbarBH}
    \bar{m}_{\ro{BH}} \sim 10^{10}\, \mathrm{gm} \left( \frac{B_0}{10^{-15}\, \mathrm{G}} \right)^{-3/2} .
\end{equation}
One can also check that for $m > \bar{m}_{\ro{BH}}$,
the scale~$H_{*}$ which is given by the first line of Eq.~(\ref{Hstar}), is comparable to or larger than the upper limit for~$H_{\ro{end}}$ given in Eq.~(\ref{eq:2.28}); hence the assumption in Eq.~(\ref{eq:HH}) breaks down. 
Therefore only the second line of the reheating bound in Eq.~\eqref{eq:perri} applies for extremal black holes, which is rewritten as
\begin{equation}
\label{eq:perriBHrh}
    F \lesssim
   10^{-18}\, \mathrm{cm}^{-2} \mathrm{sr}^{-1} \mathrm{s}^{-1} 
\left( \frac{m}{10^{10}\, \mathrm{gm}} \right)^{-1} 
 \left( \frac{T_{\mathrm{dom}}}{10^{6}\, \mathrm{GeV}} \right) 
\left( \frac{v_0}{10^{-3}} \right) .
\end{equation}
This bound applies to the mass range 
$10^{-3}\, \ro{gm} \lesssim m < \bar{m}_{\ro{BH}}$, 
given that the magnetogenesis and inflation scales satisfy the conditions in Eqs.~(\ref{eq:HH}), (\ref{eq:inf_bound}), and (\ref{eq:2.28}).
At $m > \bar{m}_{\ro{BH}}$, the bound is weaker than in the first line of Eq.~\eqref{eq:perri}
as discussed in~\cite{Kobayashi:2022qpl}, however we will not analyze this in detail.

Extremal magnetic black holes as light as $m \lesssim 10^{-3}\, \mathrm{gm}$
(i.e. $g < g_{\mathrm{min}}$)
move at nonrelativistic~$v_{\ro{H}}$ throughout the early cosmic history. The condition for such black holes not to alter the redshifting of the magnetic fields during radiation domination is Eq.~\eqref{eq:BoundDuring2}, which is now rewritten as
\begin{equation}
\label{eq:BoundDuringBH}
     F \lesssim  10^{-14}~\mathrm{cm}^{-2} \mathrm{sr}^{-1} \mathrm{s}^{-1}  
\left( \frac{m}{10^{-3}~\mathrm{gm}} \right)^{-1}
\left( \frac{v_0}{10^{-3}} \right)  .
\end{equation}
The condition from the reheating epoch has the same expression as Eq.~(\ref{eq:perriBHrh}).

\begin{figure}[!t]
  \centering
  \includegraphics[width=\textwidth]{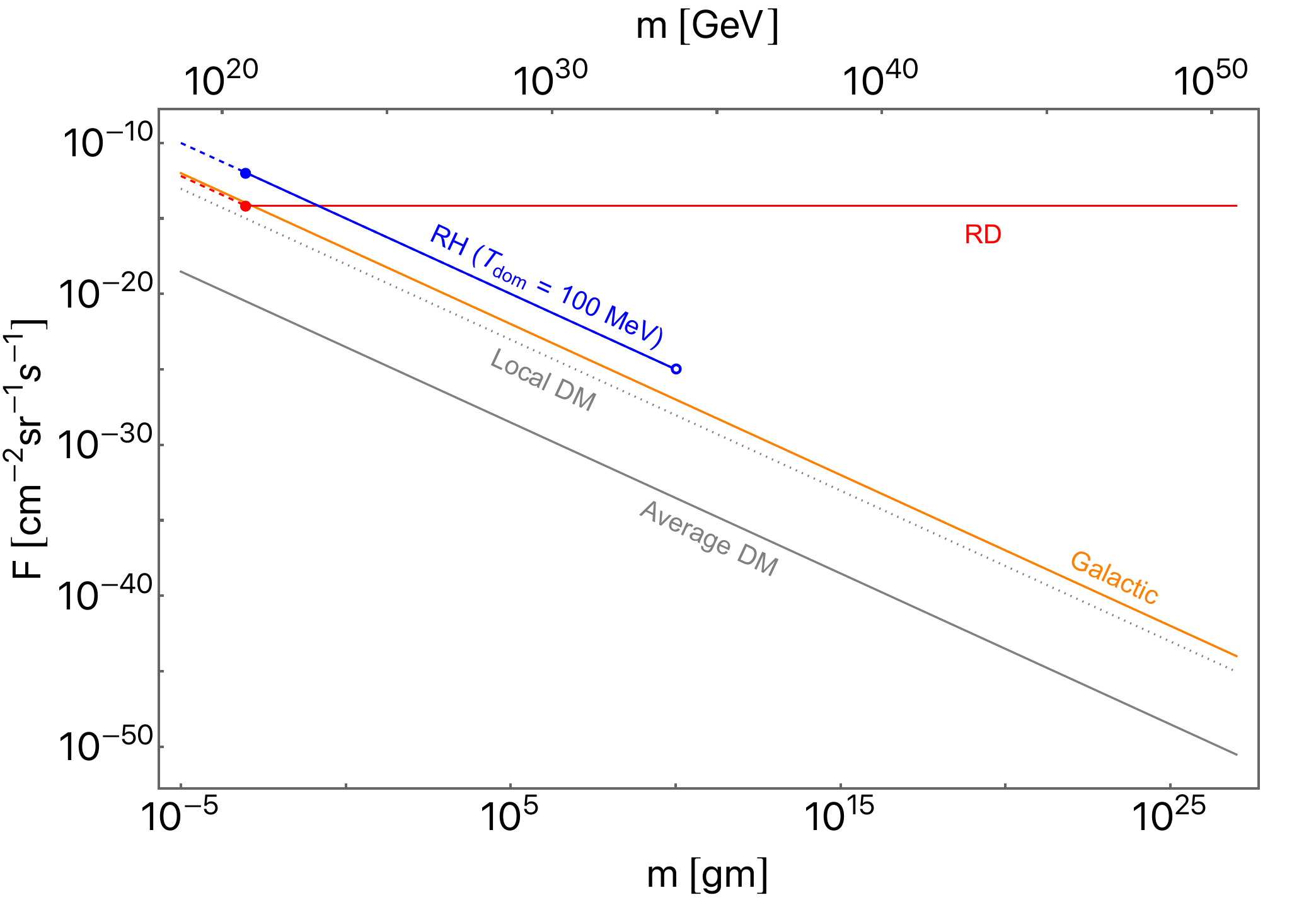}
\caption{Upper bounds on the flux of extremal magnetic black holes. 
Gray solid: abundance bound from comparison with the average dark matter density in the universe.
Gray dotted: abundance bound from comparison with the local dark matter density in our Galaxy. 
Orange: bound from Galactic magnetic fields. 
Red: bound from primordial magnetic fields during radiation domination, with present-day strength $B_0 = 10^{-15}~\mathrm{G}$.
Blue: bound from primordial magnetic fields during reheating,
with present-day strength $B_0 = 10^{-15}~\mathrm{G}$ and reheating temperature $T_{\mathrm{dom}} = 100~\mathrm{MeV}$. 
For the primordial bounds, the dotted parts of the lines show where the fields exhibit modified redshifting behaviors, instead of being dissipated.
Here we assume $v_0 = 10^{-3}$ and for the Galactic parameters $l_{\mathrm{c}} = 1~\mathrm{kpc}$, $R = 10~\mathrm{kpc}$, $\tau_{\mathrm{gen}} = 10^8~\mathrm{yr}$, and $\gamma_i - 1 = 10^{-6}$. See the text for details.
}
\label{fig:BH}
\end{figure}

\subsection{Comparison of bounds}
\label{sec:summBH}

In Figure~\ref{fig:BH} we show the upper bounds on the flux of extremal magnetic black holes. 
The solid gray line shows the cosmological abundance bound in Eq.~\eqref{eq:cosmological} with $\rho_{\mathrm{DM}}$ taken as the average dark matter density in the universe, i.e. $\rho_{\mathrm{DM}} \approx 1.3 \times 10^{-6}~\mathrm{GeV cm^{-3}}$.
The dotted gray line shows the abundance bound with $\rho_{\mathrm{DM}}$ set to the local dark matter density in the Milky Way, i.e. $\rho_{\ro{DM}} \approx 0.4 \,\ro{GeV} \, \ro{cm}^{-3}$. In both of the abundance bounds we also used $v_i = 10^{-3}$.
The orange line shows the bound in Eq.~\eqref{eq:parkerBH} from the survival of Galactic magnetic fields, with the parameters taken as $l_{\mathrm{c}} = 1~\mathrm{kpc}$, $R = 10~\mathrm{kpc}$, $\tau_{\mathrm{gen}} = 10^8~\mathrm{yr}$, and $\gamma_i - 1 = 10^{-6}$; 
this bound is independent of the Galactic field strength (as long as $B \lesssim 10^{-3}\, \ro{G}$), hence the present-day and seed Galactic fields give similar bounds.
We note that the condition in Eq.~\eqref{eq:maru-n} holds for all the values of the mass of the black holes shown in the plot, as long as $B \gtrsim 10^{-11} \, \ro{G}$.
The red line shows the bounds in Eqs.~\eqref{eq:BoundDuring} and \eqref{eq:BoundDuringBH} from primordial fields during radiation domination. 
The blue line shows the bound in Eq.~\eqref{eq:perriBHrh} from primordial fields during reheating for $T_{\mathrm{dom}} = 100\, \mathrm{MeV}$. 
For the primordial bounds we assume a present-day field strength $B_0 = 10^{-15}\, \mathrm{G}$, and velocity $v_0 = 10^{-3}$.
The filled points on the primordial bounds show where the mass limit of Eq.~\eqref{eq:BHmassLim} is saturated, and the dashed parts of the lines show where the fields exhibit modified redshifting behaviors, instead of being dissipated.
Differently from Figure~\ref{fig:Primordial}, the dashed parts of the bounds are now on the lower mass end. 
The blue open circle shows the threshold mass~$\bar{m}_{\ro{BH}}$ given in Eq.~\eqref{eq:mbarBH}.
A reheating bound also exists at $m > \bar{m}_{\ro{BH}}$, however we did not analyze this case and hence the blue line is truncated at~$\bar{m}_{\ro{BH}}$.
Let us also note that an extremal magnetic black hole with a Dirac charge $g = g_{\mathrm{D}}$ has a mass of $7.1 \times 10^{19}~\mathrm{GeV}$. 
However, lighter extremal black holes can in principle exist by absorbing minicharged monopoles.

The bound from primordial fields during radiation domination does not depend on the mass of the black holes, while the other Parker bounds become stronger for larger masses. 
Consequently, the radiation domination bound is much less constraining.
The reheating bound, even with the rather low reheating temperature chosen in the plot, is weaker than the Galactic bound; 
this is also seen in Figure~\ref{fig:boundComparison} for the mass-dependent segments of the reheating and Galactic bounds.

We also see that the abundance bound is stronger than the Galactic Parker bound, even when considering the local dark matter density.
The Parker bound can in principle be significantly improved by considering  galaxies hosting magnetic fields with coherence lengths much larger 
than the Milky Way; however it is unlikely that magnetic black holes can be clustered with such galaxies, as was discussed at the end of Section~\ref{sec:5.1}.
And if magnetic black holes cannot cluster with some galaxies, then they cannot make up all the dark matter.

\subsection{Comments on black hole-specific features}
\label{sec:monApprox}

In the above discussions we have treated extremal magnetic black holes simply as very massive monopoles with charges much larger than the Dirac charge, and ignored black hole-specific features.
However if accretion disks form around the black holes in galaxies, the interaction between the disks and the interstellar medium may affect the acceleration of black holes along the galactic fields. 
On the other hand, there has not been enough time for accretion disks to form in the early universe~\cite{Ricotti:2007au}, hence this should not affect the bounds from primordial magnetic fields. 

Extremal magnetic black holes can also be surrounded by an electroweak corona, where the value of the Higgs field varies \cite{Maldacena:2020skw, Bai:2020spd}.
The presence of electroweak coronas can also change the interaction between the black holes and the interstellar medium or the primordial plasma, modifying the Parker-type bounds. 
We leave detailed studies of these effects for the future.

\section{Conclusion}
\label{sec:concl}

We carried out a comprehensive study of the Parker-type bounds on magnetic monopoles with arbitrary charge.
We summarized the bounds from galactic magnetic fields in Section~\ref{sec:summGal}, 
and the bounds from primordial magnetic fields in Section~\ref{sec:summPrim}.
The various bounds were compared in Figure~\ref{fig:boundComparison}. 
We showed that heavy monopoles are mainly constrained by the dark matter density limit, while intermediate to low mass monopoles are mainly constrained by the Parker bounds. Among the Parker bounds, the seed galactic field bound strongly constrains monopoles with a Dirac charge, while the primordial bound from radiation domination can be the strongest for monopoles with small magnetic charges. This is because the bound from radiation domination in the low-mass regime is independent of the monopole charge, while the other Parker bounds become weaker for smaller charges.

While monopoles with a Dirac charge have to be heavier than $10^{18}\, \ro{GeV}$ to be able to cluster with our Galaxy, minicharged monopoles can cluster with much lighter masses.
For monopoles that can cluster with our Galaxy, the Parker bounds are generically less constraining than the bound from the dark matter density. Such monopoles can thus make up the entire dark matter.

We also studied extremal magnetic black holes, for which the various bounds were compared in Figure~\ref{fig:BH}. We found that extremal magnetic black holes are mainly constrained by comparison with the dark matter density.
Even stronger constraints can in principle be obtained if there exist  galaxies whose magnetic fields have coherence lengths much larger than the Milky Way.
However the large coherence lengths also lead to the acceleration of black holes up to the escape velocity within a rather short time period, 
and hence it is improbable that black holes remain clustered with such galaxies until today.
The existence of galaxies not being able to host magnetic black holes, if confirmed, would rule out the possibility of magnetic black holes as a dark matter candidate. 

Minicharged monopoles are typically connected by dark strings, whose tension is set by the mass~$\mu$ of dark photons;
moreover they appear as minicharged monopoles only at distances larger than $1 / \mu$ (see e.g. \cite{Hook:2017vyc,Hiramatsu:2021kvu}). 
In our analyses we ignored these effects, supposing that the force from the background magnetic field is stronger than the string tension, and that the field's coherence length is larger than $1 / \mu$. 
These assumptions can break down depending on the dark photon mass,
in which case our bounds can be modified.
For extremal black holes, some of the Parker bounds could be modified by the presence of accretion disks and/or the electroweak corona~\cite{Maldacena:2020skw, Bai:2020spd}.
We leave detailed considerations of these effects for future analysis.

\acknowledgments

We thank Kyrylo Bondarenko, Michele Doro, Rajeev Kumar Jain, Maurizio Spurio, and Piero Ullio for helpful discussions.
T.K. acknowledges support from the INFN program on Theoretical Astroparticle Physics (TAsP), and JSPS KAKENHI (Grant No.~JP22K03595).

\appendix

\section{Acceleration of monopoles in galactic magnetic fields}
\label{app:dynamics}

Here we study monopole dynamics in galactic magnetic fields. We divide the magnetic field region of the galaxy into cells of uniform field, and analyze the acceleration of monopoles as they pass through multiple cells. 

The equation of motion of a monopole passing through the $N$th cell
with uniform magnetic field~$\bd{B}_N$ is
\begin{equation}
 m \frac{ d  }{dt} (\gamma \bd{v} ) = g \bd{B}_N,
\end{equation}
where $\gamma = 1/\sqrt{1-v^2}$ and  $v = \abs{\bd{v}}$.
($g$ denotes the amplitude of the magnetic charge, i.e. $g > 0$, and thus the monopole here has a positive charge. However the discussion in this appendix can be applied to negatively charged monopoles by replacing $\bd{v} \to - \bd{v}$.)
By integrating the equation, one obtains the change in the monopole's Lorentz factor in the $N$th cell as,
\begin{equation}
 \gamma_{N}^2 -  \gamma_{N-1}^2 = 
\left( \frac{g B \tau_{N} }{m}  \right)^2
+ \frac{2 g \bd{B}_N \cdot \bd{v}_{N-1} \gamma_{N-1} \tau_N}{m}.
\label{eq:maru-1}
\end{equation}
Here $\tau_N$ denotes the time it takes for the monopole to pass through the $N$th cell, and $\gamma_N$ is the Lorentz factor when the monopole exits the $N$th cell and simultaneously enters the $(N+1)$th cell;
the same notation is used for the velocity~$\bd{v}_N$.
For $N=1$, then $\gamma_{N-1}$ and $\bd{v}_{N-1}$ in the equation are replaced by the initial Lorentz factor $\gamma_{i}$ and velocity $\bd{v}_{i}$ upon entering the first cell.

We take all cells to have the same size~$l_{\ro{c}}$ and field strength, i.e. $B = \abs{\bd{B}_N}$ for all~$N$.
Thus the kinetic energy of a monopole changes within each cell by at most $\sim g B l_{\ro{c}}$.
If the kinetic energy is initially large such that $m (\gamma_i - 1) \gg g B l_{\ro{c}}$, then the energy barely changes in the first cell.
On the other hand if $m (\gamma_i - 1) \ll g B l_{\ro{c}}$, the monopole is quickly accelerated so that upon exiting the first cell its energy reaches $m (\gamma_1 - 1) \simeq g B l_{\ro{c}}$, and thereafter the energy does not change much within each cell. 
Hence independently of $\gamma_i$, we can write the crossing time for the second cell onward as\footnote{The exact value of $\tau_{N}$ also depends on the shape of the cell and the incident angle, however the expression (\ref{eq:tau_N}) is good enough for our purpose of obtaining an order-of-magnitude estimate of the average energy gain.}
\begin{equation}
 \tau_N \sim \frac{l_{\ro{c}}}{v_{N-1}}
\quad \ro{for} \,\, 
N \geq 2.
\label{eq:tau_N}
\end{equation}

Let us consider nonrelativistic monopoles for the moment.
Then (\ref{eq:maru-1}) at $N \geq 2$ can be rewritten using (\ref{eq:tau_N}) as,
\begin{equation}
 v_{N}^2 - v_{N-1}^2 = \frac{v_{\ro{mag}}^4}{4 v_{N-1}^2}
+ v_{\ro{mag}}^2 
\hat{\bd{B}}_N \cdot \hat{\bd{v}}_{N-1}.
\label{eq:A.4}
\end{equation}
Here a hat denotes a unit vector:
$\hat{\bd{B}}_N \equiv \bd{B}_N / B$ and 
$\hat{\bd{v}}_N \equiv \bd{v}_N / v_N$.
We also introduced
\begin{equation}
v_{\ro{mag}}  \equiv \sqrt{ \frac{2 g B l_{\ro{c}}}{m} },
\end{equation}
which corresponds to the velocity a monopole initially at rest obtains after passing through a single cell.
From the discussions above (\ref{eq:tau_N}) it follows that
$v_1 \gtrsim v_{\ro{mag}}$ for general~$v_i$.

Supposing for simplicity that the direction of the magnetic field is uncorrelated from one cell to the next, the second term in the right-hand side of (\ref{eq:A.4}) sources a random walk of~$v^2$
in each cell. As we are interested in the mean behavior of the monopoles, let us ignore this term for now.
Then we obtain a recurrence relation of the form\footnote{Here we are also roughly approximating the mean $\langle 1/v^{2}_{N-1} \rangle$ by $1/\langle v^{2}_{N-1} \rangle$.}
\begin{equation}
 \beta_{N} - \beta_{N-1} = \frac{1}{\beta_{N-1}},
\label{eq:beta-rec}
\end{equation}
where $\beta_N \equiv 2 v_N^2 / v_{\ro{mag}}^2$.
Since $\beta_1 \gtrsim 1$, this recurrence relation has an approximate solution,
\begin{equation}
 \beta_N \simeq \sqrt{ \beta_1^2 + 2 (N-1) } .
\label{eq:beta-approx}
\end{equation}
Hence the exit velocity from the $N$th cell is obtained as
\begin{equation}
   v_N^2 \simeq 
\sqrt{ v_1^4 + \frac{N-1}{2} v_{\ro{mag}}^4 }.
\label{eq:A.8}
\end{equation}

If $v_i \gtrsim v_{\ro{mag}}$, the discussions from (\ref{eq:tau_N}) onward apply also to $N = 1$, then one can make the replacements 
$v_1 \to v_i$ and $N-1 \to N$ in the right-hand side of (\ref{eq:A.8}).
On the other hand if $v_i \ll v_{\ro{mag}}$, then $v_1 \simeq  v_{\ro{mag}}$ and (\ref{eq:A.8}) becomes
$v_N^2 \simeq \sqrt{(N+1)/2} \, v_{\ro{mag}}^2$.
In both cases, (\ref{eq:A.8}) can be rewritten at the order-of-magnitude level as
\begin{equation}
   v_N^2 \sim \sqrt{ v_i^4 + \frac{N}{2} v_{\ro{mag}}^4 } .
\end{equation}
In particular, the net change in the velocity squared in the limit of 
small and large $N$ takes the forms,
\begin{equation}
\Delta v_N^2 = v_N^2  - v_i^2 \sim 
  \begin{dcases}
\frac{N }{4} \frac{v_{\ro{mag}}^4}{v_i^2}
  & \mathrm{for}\, \, \,  
N \ll 8 \left( \frac{v_i}{v_{\ro{mag}}} \right)^4 ,
 \\
\sqrt{\frac{N}{2}} \, v_{\ro{mag}}^2
  & \mathrm{for}\, \, \,  
N \gg 8 \left( \frac{v_i}{v_{\ro{mag}}} \right)^4 .
 \end{dcases}
\label{eq:A.10}
\end{equation}
In the first line the acceleration is tiny such that
$\Delta v_N \lesssim v_i$; this regime exists only if $v_i \gtrsim v_{\ro{mag}}$.
Eventually the monopole is accelerated as in the second line, 
where $\Delta v_N \gtrsim v_i$.

Let us discuss the second term in (\ref{eq:A.4}) which we have been ignoring. This sources a random walk behavior of $\Delta v^2$ in each cell
with step size~$\leq v_{\ro{mag}}^2$, which after $N$~cells yields a root-mean-square distance of order~$\sqrt{N} v_{\ro{mag}}^2$.
Now, consider $p$~number of monopoles with initial velocity~$v_i$,
each passing through $N$~cells in different parts of the galaxy.
From the central limit theorem, the distribution of the average of $\Delta v_N^2$ with large enough~$p$ is approximated by a normal distribution with mean (\ref{eq:A.10}) and standard deviation of
\begin{equation}
 \sigma \sim \sqrt{\frac{N}{p}} \, v_{\ro{mag}}^2.
\label{eq:StDe}
\end{equation}
The expression (\ref{eq:A.10}) describes well the average behavior for the set of monopoles if it is much larger than~$\sigma$.
For this, the second line of (\ref{eq:A.10}) requires only $p \gg 1$, while the first line requires
\begin{equation}
 p N  \gg 16 \left( \frac{v_i}{v_{\ro{mag}}} \right)^4.
\label{eq:A.12}
\end{equation}

For relativistic monopoles ($v_N \simeq 1$), the mean recurrence relation becomes
\begin{equation}
 \gamma_N^2 -  \gamma_{N-1}^2 = \left( \frac{g B l_{\ro{c}}}{m} \right)^2,
\end{equation}
which yields
\begin{equation}
 \gamma_N = \sqrt{ \gamma_1^2 + (N-1) \left(\frac{g B l_{\ro{c}}}{m}\right)^2 }.
\end{equation}
By following a similar analysis as for nonrelativistic monopoles, one arrives at results that match at the order-of-magnitude level with (\ref{eq:A.10}) and (\ref{eq:A.12}), with $v^2$ replaced by $2 (\gamma - 1)$.

In summary, for both nonrelativistic and relativistic monopoles, the average energy gain after passing through $N$~cells takes the form
\begin{equation}
\Delta E_N = m (\gamma_N - \gamma_i) \sim 
  \begin{dcases}
\frac{N}{4} 
\frac{(g B l_{\ro{c}})^2}{m (\gamma_i - 1)}
  & \mathrm{for}\, \, \,  
N \ll 8 \left( \frac{m (\gamma_i - 1)}{g B l_{\ro{c}}} \right)^2 ,
 \\
\sqrt{\frac{N}{2}} \, g B l_{\ro{c}}
  & \mathrm{for}\, \, \,  
N \gg 8 \left( \frac{m (\gamma_i - 1)}{g B l_{\ro{c}}} \right)^2 .
 \end{dcases}
\end{equation}
For this to describe well the average behavior of a set of monopoles, the first line requires a sufficiently large number of monopoles~$p$ such that
\begin{equation}
 p N \gg 16 \left( \dfrac{m (\gamma_i - 1)}{g B l_{\ro{c}}} \right)^2 ,
\end{equation}
while the second line requires only $p \gg 1$.

\end{document}